\documentclass[manuscript , screen ]{acmart}

\AtBeginDocument{%
  \providecommand\BibTeX{{%
    \normalfont B\kern-0.5em{\scshape i\kern-0.25em b}\kern-0.8em\TeX}}}

\setcopyright{acmcopyright}
\copyrightyear{2018}
\acmYear{2018}
\usepackage{graphicx}
\usepackage{color, soul}  % for underline color 
\usepackage{colortbl}
\usepackage{enumitem}
\usepackage{amsmath}
\usepackage[caption=false,font=footnotesize,labelformat=simple]{subfig}
\begin{document}

\title{Security Smells in Ansible and Chef Scripts: A Replication Study} 

\author{Akond Rahman}
\orcid{0000-0002-5056-757X}
\affiliation{%
  \institution{Tennessee Technological University}
  \streetaddress{1 William Jones Drive}
  \city{Cookeville}
  \state{Tennessee}
  \country{USA}
}
\email{arahman@tntech.edu}

\author{Md Rayhanur Rahman}
\affiliation{%
  \institution{NC State University}
  \streetaddress{890 Oval Drive}
  \city{Raleigh}
  \state{North Carolina}
  \country{USA}
}
\email{mrahman@ncsu.edu}

\author{Chris Parnin}
\affiliation{%
  \institution{NC State University}
  \streetaddress{890 Oval Drive}
  \city{Raleigh}
  \state{North Carolina}
  \country{USA}
}
\email{cjparnin@ncsu.edu}

\author{Laurie Williams}
\affiliation{%
  \institution{NC State University}
  \streetaddress{890 Oval Drive}
  \city{Raleigh}
  \state{North Carolina}
  \country{USA}
}
\email{lawilli3@ncsu.edu}

\renewcommand{\shortauthors}{Rahman et al.}

\begin{abstract}
\textbf{Context:} Security smells are recurring coding patterns that are indicative of security weakness, and require further inspection. As infrastructure as code (IaC) scripts, such as Ansible and Chef scripts, are used to provision cloud-based servers and systems at scale, security smells in IaC scripts could be used to enable malicious users to exploit vulnerabilities in the provisioned systems. \textbf{Goal:} \textit{The goal of this paper is to help practitioners avoid insecure coding practices while developing infrastructure as code scripts through an empirical study of security smells in Ansible and Chef scripts.} \textbf{Methodology:} We conduct a replication study where we apply qualitative analysis with 1,956 IaC scripts to identify security smells for IaC scripts written in two languages: Ansible and Chef. We construct a static analysis tool called \textbf{S}ecurity \textbf{L}inter for \textbf{A}nsible and \textbf{C}hef scripts (SLAC) to automatically identify security smells in 50,323 scripts collected from 813 open source software repositories. We also submit bug reports for 1,000 randomly-selected smell occurrences. \textbf{Results:} We identify two security smells not reported in prior work: missing default in case statement and no integrity check. By applying SLAC we identify 46,600 occurrences of security smells that include 7,849 hard-coded passwords. We observe agreement for 65 of the responded 94 bug reports, which suggests the relevance of security smells for Ansible and Chef scripts amongst practitioners. \textbf{Conclusion:} We observe security smells to be prevalent in Ansible and Chef scripts, similar to that of the Puppet scripts. We recommend practitioners to rigorously inspect the presence of the identified security smells in Ansible and Chef scripts using (i) code review, and (ii) static analysis tools. 
\\ 
{\color{red} The paper is accepted at the journal of ACM Transactions on Software Engineering and Methodology (TOSEM) on June 20, 2020. }
\end{abstract}




\maketitle

%%%%%%%%%%%%%%%%%%%%%%%%%%%%%%%%%%%%%%%%%%%%%%%%% INTRO %%%%%%%%%%%%%%%%%%%%%%%%%%%%%%%%%%%%%%%%%%%%%%%%%%%%%%%
%%%%%%%%%%%%%%%%%%%%%%%%%%%%%%%%%%%%%%%%%%%%%%%%%%%%%%%%%%%%%%%%%%%%%%%%%%%%%%%%%%%%%%%%%%%%%%%%%%%%%%%%%%%%%%%

\section{Introduction}
\label{intro} 

Infrastructure as code (IaC) is the practice of using automated scripting to provision and configure their development environment and servers at scale~\cite{Humble:2010:CD}. Similar to software source code, recommended software engineering practices, such as version control and testing are expected to be applied to implement the practice of IaC. IaC tool vendors, such as Ansible~\footnote{https://www.ansible.com/} and Chef~\footnote{https://www.chef.io/chef/} provide programming utilities to implement the practice of IaC. The use of IaC scripts has resulted in benefits for information technology (IT) organizations. For example, the use of IaC scripts helped the National Aeronautics and Space Administration (NASA) to reduce its multi-day patching process to 45 minutes~\cite{nasa:iac}. Using IaC scripts application deployment time for Borsa Istanbul, Turkey's stock exchange, reduced from $\mathtt{\sim}$10 days to an hour~\cite{borsa:pup}. With IaC scripts Ambit Energy increased their deployment frequency by a factor of 1,200~\cite{ambit:pup}.  The Enterprise Strategy Group surveyed practitioners and reported the use of IaC scripts to help IT organizations gain 210\% in time savings and 97\% in cost savings on average~\cite{ESG:iac}.

Despite reported benefits, IaC scripts can be susceptible to security weakness. In our recent work, we identified security smells for Puppet scripts~\cite{me:icse2019:slic}. Security smells are recurring coding patterns that are indicative of security weakness, and requires further inspection~\cite{me:icse2019:slic}. We identified 21,201 occurrences of seven security smells that include 1,326 occurrences of hard-coded passwords in 15,232 Puppet scripts. Our prior research showed relevance of the identified security smells amongst practitioners as well: from 212 responses we observe practitioners to agree with 148 occurrences.   

IT organizations may use other languages, such as Ansible, Chef, and Terraform~\footnote{https://www.terraform.io/}, for which our previous categorization of security smells reported in prior work~\cite{me:icse2019:slic} may not hold. A replication of our prior work for other languages, such as Ansible and Chef, may have value for practitioners as well as for research as we study the generalizability and robustness of IaC security smells in a larger variety of contexts. A 2019 survey with 786 practitioners reported Ansible as the most popular language to implement IaC, followed by Chef~\footnote{https://info.flexerasoftware.com/SLO-WP-State-of-the-Cloud-2019}~\footnote{https://www.techrepublic.com/article/ansible-overtakes-chef-and-puppet-as-the-top-cloud-configuration-management-tool/}. As usage of Ansible and Chef is getting increasingly popular amongst practitioners, identification of security smells could have relevance to practitioners in mitigating insecure coding practices in IaC.   

\begin{figure}[]
\centering
\includegraphics[scale=0.95]{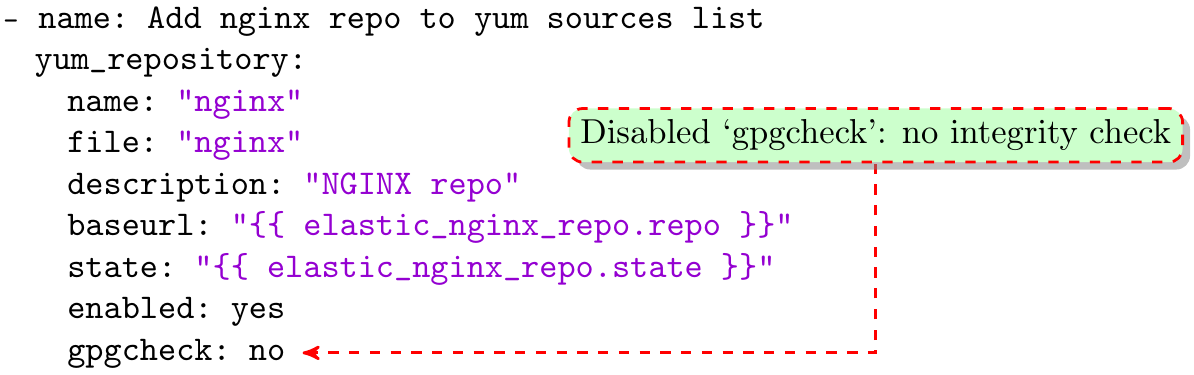}
\caption{An example Ansible script where integrity check is not specified.}
\label{fig-intro-ansi}
\end{figure}

Our prior research~\cite{me:icse2019:slic} is not exhaustive and may not capture security smells that exist for other languages. Let us consider Figure~\ref{fig-intro-ansi} in this regard. In Figure~\ref{fig-intro-ansi}, we present an actual Ansible code snippet downloaded from an open source software (OSS) repository~\footnote{https://git.openstack.org/cgit/openstack/openstack-ansible-ops/}. In the code snippet, we observe the `gpgcheck' parameter is assigned `no', indicating while downloading the `nginx' package, the `yum' package manager will not check the contents of the downloaded package~\footnote{https://docs.ansible.com/ansible/2.3/yum\_repository\_module.html}. Not checking the content of a downloaded package is related to a security weakness called `Download of Code Without Integrity Check (CWE-494)~\footnote{https://cwe.mitre.org/data/definitions/494.html}'. According to Common Weakness Enumeration (CWE), not specifying integrity check may help malicious users to ``\textit{execute attacker-controlled commands, read or modify sensitive resources, or prevent the software from functioning correctly for legitimate users}''. 

%The above-mentioned example shows security smells exist for other IaC-related languages not identified in our prior work~\cite{me:icse2019:slic}. 

Existence and persistence of security smells similar to Figure~\ref{fig-intro-ansi} in IaC scripts provide attackers opportunities to attack the provisioned system. We hypothesize through a replication~\cite{schmidt2009:replication} of our prior work, we can systematically identify security smells for other languages namely, Ansible and Chef.  

\textit{The goal of this paper is to help practitioners avoid insecure coding practices while developing infrastructure as code scripts through an empirical study of security smells in Ansible and Chef scripts.} 
 
We answer the following research questions: 

\begin{itemize} 
\item{\textbf{RQ1}: What security smells occur in Ansible and Chef scripts? }
\item{\textbf{RQ2}: How frequently do security smells occur for Ansible and Chef scripts? }
\item{\textbf{RQ3}: How do practitioners perceive the identified security smell occurrences for Ansible and Chef scripts? }
\end{itemize}

We build on prior research~\cite{me:icse2019:slic} related to security smells for IaC scripts on Puppet, and investigate what security smells for two languages used to implement the practice of IaC, namely Ansible and Chef. We conduct a differentiated replication~\cite{juristo2010replication}~\cite{Krein2010:replication} of our prior work~\cite{me:icse2019:slic}, where we use a experimental setup different to our prior work using Ansible and Chef scripts. We apply qualitative analysis~\cite{wohlin:ese} on 1,101 Ansible scripts and 855 Chef scripts to determine security smells. Next, we construct a static analysis tool called \textbf{S}ecurity \textbf{L}inter for \textbf{A}nsible and \textbf{C}hef scripts (\textbf{SLAC})~\cite{me:icse2019:slic} to automatically identify the occurrence of these security smells in 14,253 Ansible and 36,070 Chef scripts collected by respectively, mining 365 and 448 OSS repositories. We calculate smell density for each type of security smell in the collected IaC scripts. We submit bug reports for 1,000 randomly-selected smell occurrences for Ansible and Chef to assess the relevance of the identified security smells.    

\textbf{Contributions}: Compared to our prior research~\cite{me:icse2019:slic} in which we reported findings specific to Puppet, we make the following additional contributions:

\begin{itemize} 
\item{A list of security smells for Ansible and Chef scripts that include two categories not reported in prior work\cite{me:icse2019:slic}}; 
\item{An evaluation of security smell frequency occuring in Ansible and Chef scripts. As a result of this evaluation, we have created a benchmark of how frequently security smells appear for Ansible and Chef which was missing for the two languages. The frequency of identified security smells for Ansible and Chef scripts can be used as a measuring stick by practitioners and researchers alike};
\item{A detailed discussion on how practitioner responses from bug reports can drive actionable detection and repair of Ansible and Chef security smells. In our prior work, we did not discuss how the practitioner’s responses in bug reports can guide tools for actionable detection and repair}; 
\item{An empirically-validated tool (SLAC) that automatically detects occurrences of security smells for Ansible and Chef scripts. The tool that we constructed as part of prior work will not work for Ansible and Chef scripts. The `Parser' component of SLAC is different to that of `SLIC' that we built in our prior work. The `Rule Engine' component of SLAC is different to that of SLIC~\cite{me:icse2019:slic}, as unlike Puppet, which uses attributes, Ansible and Chef respectively, uses `Keys' and `Properties'}; and 
\item{A discussion of differences between the three IaC languages: Ansible, Chef, and Puppet. In our prior work, we provided background on Puppet scripts only, and did not discuss the differences between Ansible, Chef, and Puppet}. 
\end{itemize}

% full path of code snippet: https://git.openstack.org/cgit/openstack/openstack-ansible-ops/tree/elk_metrics_6x/roles/elastic_repositories/tasks/elastic_yum_repos.yml 

We organize the rest of the paper as following: we provide background information with related work discussion in Section~\ref{bg-rel}. We describe the methodology and the definitions of identified security smells in Section~\ref{security-smells}. We describe the methodology to construct and evaluate SLAC in Section~\ref{slic}. In Section~\ref{empirical}, we describe the methodology for our empirical study. We report our findings in Section~\ref{results}, followed by a discussion in Section~\ref{discussion}. We describe limitations in Section~\ref{threats}, and conclude our paper in Section~\ref{conclusion}.

%%%%%%%%%%%%%%%%%%%%%%%%%%%%%%%%%%%%%%%%%%%%%%%%% BACKGROUND AND RELATED WORK %%%%%%%%%%%%%%%%%%%%%%%%%%%%%%%%%%
%%%%%%%%%%%%%%%%%%%%%%%%%%%%%%%%%%%%%%%%%%%%%%%%%%%%%%%%%%%%%%%%%%%%%%%%%%%%%%%%%%%%%%%%%%%%%%%%%%%%%%%%%%%%%%%%
 
\section{Background and Related Work}
\label{bg-rel} 

We provide background information with related work discussion in this section.

\subsection{Background}
\label{bg}

In this section we provide background on Ansible and Chef scripts, along with CWE, as we use CWE to validate our qualitative process described in Section~\ref{meth-cgt}.   

\subsubsection{Ansible and Chef Scripts} 
\label{bg-iac}

\begin{figure}[t]
\centering
\includegraphics[scale=0.95]{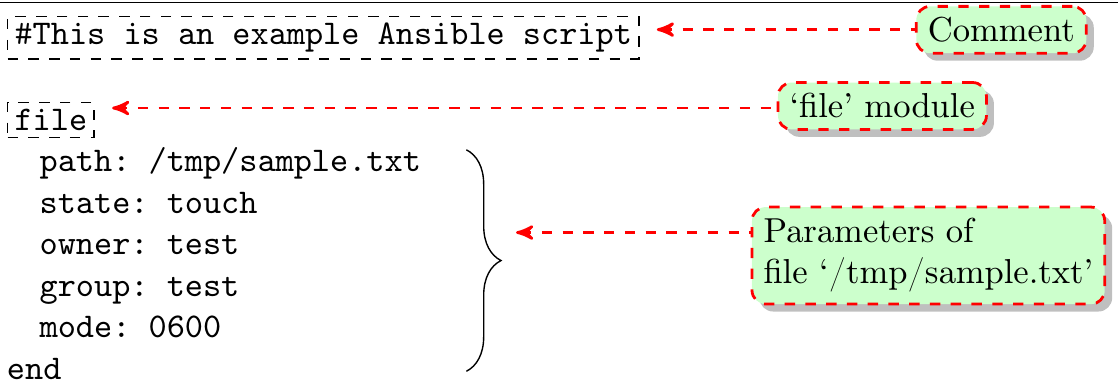}
\caption{Annotation of an example Ansible script.}
\label{fig-bg-ansi}
\end{figure}

We provide a brief background on Ansible and Chef scripts, which is relevant to conduct our empirical study. Both, Ansible and Chef provide multiple libraries to manage infrastructure and system configurations. In the case of Ansible, developers can manage configurations using `playbooks', which uses YAML files to manage configurations. For example, as shown in Figure~\ref{fig-bg-ansi}, an empty file `/tmp/sample.txt' is created using the `file' module provided by Ansible. The properties of the file such as, path, owner, and group can also be specified. The `state' property provides options to create an empty file using the `touch' value. 

\begin{figure}[t]
\centering
\includegraphics[scale=0.95]{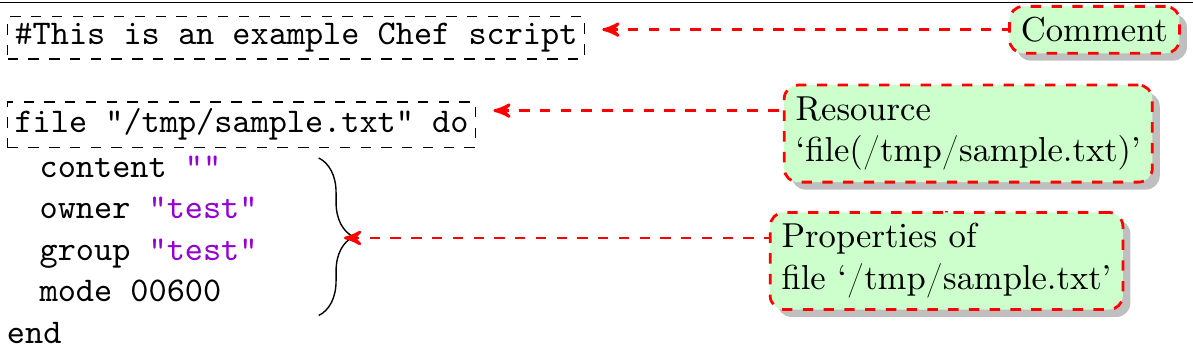}
\caption{Annotation of an example Chef script.}
\label{fig-bg-chef}
\end{figure}

In the case of Chef, configurations are specified using `recipes', which are domain-specific Ruby scripts. Dedicated libraries are also available to maintain certain configurations. As shown in Figure~\ref{fig-bg-chef}, using the `file' resource, an empty file `/var/sample.txt' is created. The `content' property is used to specify the content of the file is empty. 

\subsubsection{Differences between Ansible, Chef, and Puppet}
\label{bg-differences} 

The three languages are different to each other with respect to execution order, perceived codebase maintenance, requiring additional agent software installation, style and syntax. We discuss each of these differences below and also present a summary of the differences in Table~\ref{tab-bg-diffs}.   

\begin{itemize}

    \item{\textit{Construction}: Ansible is created with Python, whereas Chef and Puppet are created using Ruby. 
    }    

    \item{\textit{Execution order}: For procedural configuration languages, such as Ansible and Chef, understanding of the order in which tasks are executed is important because specifying a different order might provision the desired infrastructure differently. On the other hand, for Puppet, the current code state provides a clear view of what will be the configurations of the provisioned infrastructure.   
    }
    \item{\textit{Perceived codebase maintenance}: Practitioners~\cite{bg_diff} perceive Ansible and Chef code bases to be large and incur more maintenance overhead, as previously written Ansible and Chef code might be obsolete after a certain period of time. The state of the provisioned infrastructure might change constantly, and code written a week ago might become unusable, and practitioners have to write more code. Unlike Ansible and Chef, Puppet code is a direct reflection of the current state of the provisioned infrastructure, and practitioner might not need to write new code to be consistent with the current state of the provisioned infrastructure.   
    }
    \item{\textit{Requiring additional agent software}: For Chef and Puppet users installation of additional agent software is required for each of the servers that the practitioner wants to configure~\cite{chef-doc}~\cite{ansi-doc}. Typically, the agents run as background services and executes necessary updates to the provision infrastructure when needed. Practitioners~\cite{bg_diff} perceive use of agent software to have limitations related to maintenance. For example, if a defect occurs, then the practitioner needs to troubleshoot the scripts, the installed agents, as well as the communication amongst the installed agents. For Ansible, installation of additional agent software is not required.  
    }    
    \item{\textit{Style}: Ansible and Chef scripts are developed using a procedural style. Practitioners write Ansible and Chef scripts in a step-by-step manner so that the desired end state is reached. Unlike Ansible and Chef, Puppet is developed using a declarative style, where the desired state is specified first, and the Puppet tool itself is responsible to reach the desired state. In both cases, the desired state refers to the state of the provisioned computing infrastructure. For example, in the case of Figures~\ref{fig-bg-ansi} and~\ref{fig-bg-chef}, the desired state is to create an empty text file.   
    }
    \item{\textit{Syntax}: Ansible, Chef, and Puppet respectively use YAML, Ruby, and Puppet domain specific language (DSL) as their syntax. The differences in the syntax of the programming languages determine the expressiveness for the programming languages. For example, practitioners~\cite{bg_diff} have reported that declarative languages, such as Ansible and Chef can be limiting to conduct certain DevOps-related tasks, such as gradual rollouts and zero-downtime deployment. 
    }

\end{itemize}

%%%%%%%%%%%%%%%%%%%%%%%%%% Rebuttal ::: Differences %%%%%%%%%%%%%
\begin{table}[htb]
\centering
\caption{Summary of Differences between Ansible, Chef, and Puppet Scripts} 
\label{tab-bg-diffs}
\begin{tabular}{ p{1.5cm} p{1.5cm} p{2cm} p{1cm} p{1.5cm} p{1.5cm} p{1.5cm} }
\hline
\textbf{Lang.} & \textbf{Execution Order} & \textbf{Perceived Maint.} & \textbf{Add. agent} & \textbf{Style} & \textbf{Syntax} & \textbf{Construction} \\ 
\hline
Ansible & Provisioning dependent on ordering of code   & High  & No   & Declarative  & YAML       & Python\\
Chef    & Provisioning dependent on ordering of code   & High  & Yes  & Declarative  & Ruby       & Ruby\\
Puppet  & Provisioning independent on ordering of code & Low   & Yes  & Procedural   & Puppet DSL & Ruby\\
\hline
\end{tabular}
\end{table}

%%%%%%%%%%%%%%%%%%%%%%%%%% Rebuttal ::: Differences %%%%%%%%%%%%%

\subsubsection{Common Weakness Enumeration (CWE)} 
\label{bg-cwe}

CWE is a community-driven database for software security weaknesses and vulnerabilities~\cite{cwe:mitre}. The goal of creating this database is to understand security weaknesses in software, create automated tools so that security weaknesses in software can be automatically identified and repaired, and create a common baseline standard for security weakness identification, mitigation, and prevention efforts~\cite{cwe:mitre}. The database is owned by the MITRE Corporation, with support from US-CERT and the National CyberSecurity Division of the United States Department of Homeland Security~\cite{cwe:mitre}. 

\subsubsection{Differentiated Replication in Software Engineering} 
\label{bg-replication} 

We conduct a differentiated replication~\cite{Krein2010:replication} of our prior work~\cite{me:icse2019:slic}. Krein and Knutson~\cite{Krein2010:replication} constructed a replication taxonomy for software engineering research. Their taxonomy included four categories of replication, namely, strict replication, differentiated replication, dependent replication, and independent replication. In strict replication, protocols of a prior research study is strictly followed as possible. In differentiated replication, the protocol of the prior research study is intentionally altered by the researchers. Dependent replication refers to research studies that is designed with reference to one or more prior research studies. Independent replication answers the same research questions as a prior research study, but it conducted without knowledge of, or deference to the prior research study. Our research paper focuses on Ansible and Chef scripts, which necessitates alteration in the study design of our prior research paper on security smells~\cite{me:icse2019:slic}.      

%%%%%%%%%%%%%%%%%%%%%%%%%%%%%%%%%%%%%%%%%%%%% Related Work %%%%%%%%%%%%
%%%%%%%%%%%%%%%%%%%%%%%%%%%%%%%%%%%%%%%%%%%%%%%%%%%%%%%%%%%%%%%%%%%%%%%

\subsection{Related Work}
\label{rel}

For IaC scripts, we observe a lack of studies that investigate coding practices with security consequences. For example, Sharma et al.~\cite{SharmaPuppet2016}, Schwarz~\cite{undergrad:thesis:iac}, and Bent et al.~\cite{Bent:Saner2018:Puppet}, in separate studies investigated code maintainability aspects of Chef and Puppet scripts. Jiang and Adams ~\cite{JiangAdamsMSR2015} investigated the co-evolution of IaC scripts and other software artifacts, such as build files and source code. Rahman and Williams~\cite{me:icst2018:iac} characterized defective IaC scripts using text mining and created prediction models using text feature metrics. Rahman et al.~\cite{Rahman:RCOSE17} surveyed practitioners to investigate which factors influence usage of IaC tools. Rahman et al.~\cite{me:ist:sms:2018} conducted a systematic mapping study with 32 IaC-related publications and observed lack in security-related research in the domain of IaC. Rahman and Williams~\cite{me:ist2019:code:properties} identified 10 code properties in IaC scripts that show correlation with defective IaC scripts. Hanappi et al.~\cite{Hanappi:2016:pupp:converge} investigated how convergence of IaC scripts can be automatically tested, and proposed an automated model-based test framework. Rahman et al.~\cite{Rahman:2020:ACID} also constructed a defect taxonomy for IaC scripts that included eight defect categories. In another work Rahman et al.~\cite{me:emse:ap:iac} identified five development anti-patterns for IaC scripts. In this paper we build upon the research conducted by Rahman et al.~\cite{me:icse2019:slic}'s research, which identified seven types of security smells that are indicative of security weaknesses in IaC scripts. They identified 21,201 occurrences of security smells that include 1,326 occurrences of hard-coded passwords. The three languages are different to each other with respect to execution order, perceived codebase maintenance, requiring additional agent software installation, style, and syntax. Differences in IaC languages along with the need to advance the science of IaC script quality motivate us to conduct our research. We replicate Rahman et al.~\cite{me:icse2019:slic}'s research for Ansible and Chef scripts.

%%%%%%%%%%%%%%%%%%%%%%%%%%%%%%%%%%%%%%%%%%%%%%%%% SECURITY SMELLS %%%%%%%%%%%%%%%%%%%%%%%%%%%%%%%%%%%%%%%%%%%%%
%%%%%%%%%%%%%%%%%%%%%%%%%%%%%%%%%%%%%%%%%%%%%%%%%%%%%%%%%%%%%%%%%%%%%%%%%%%%%%%%%%%%%%%%%%%%%%%%%%%%%%%%%%%%%%%

\section{Security Smells}
\label{security-smells} 

A code smell is a recurrent coding pattern that is indicative of potential maintenance problems~\cite{fowler1999refactoring}. A code smell may not always have bad consequences, but still deserves attention, as a code smell may be an indicator of a problem~\cite{fowler1999refactoring}. Our paper focuses on identifying security smells. Security smells are recurring coding patterns that are indicative of security weakness, and require further inspection~\cite{me:icse2019:slic}. 

We conduct a differentiated replication of our prior research, where we alter the research questions and methodology for Puppet scripts and apply the methodology for Ansible and Chef scripts. We exclude the analysis of lifetime because before quantifying the lifetime of security smells, we wanted to understand (i) what categories of security smells exist, (ii) if security smells are frequent, and (iii) if the identified security smells have relevance to practitioners. Without establishing the groundwork that addresses all these factors, lifetime analysis would not have been relevant for practitioners. 

We describe the methodology to derive security smells in IaC scripts, followed by the definitions and examples for the identified security smells.   

\subsection{RQ1: What security smells occur in Ansible and Chef scripts? }
\label{meth-cgt} 

\hspace{0.3cm} \textbf{Data Collection}: We collect a set of Ansible and Chef scripts to determine security smells for each language. We collect 1,101 Ansible scripts that we use to determine the security smells from 16 OSS repositories maintained by Openstack. We collect 855 Chef scripts from 10 repositories maintained by Openstack. We select Openstack as Openstack provides utilities related to cloud computing and have made their source code online. Our assumption is that by collecting Ansible and Chef scripts from the repositories we will be able to obtain a sufficient amount of Ansible and Chef scripts to perform qualitative analysis. We download these repositories on Nov 11, 2018. As of November 2018, the Openstack organization made 1,253 repositories available. Of these 1,253 repositories we collect repositories for which 11\% of all files are IaC scripts. We apply this criterion because we wanted to collect a large collection of Ansible and Chef scripts so that we have sufficient amount Ansible and Chef code to investigate. All these repositories are hosted on Openstack's public repository browser~\footnote{https://git.openstack.org/cgit}, and not on GitHub. 

We provide summary statistics in Table~\ref{tab-rq1-summary-rebuttal} of the 16 Ansible and 10 Chef OSS repositories. The `IaC Cnt.' and `IaC Size' respectively, presents the total count of IaC scripts and total size of all collected IaC scripts as measured by lines of code.

%%%%%%%%%%%%%%%%%%%%%%%%%% Rebuttal ::: Summary ::: RQ1 %%%%%%%%%%%%%

\begin{table}[htb]
\centering
\caption{Summary Statistics of the Collected Repositories Used in RQ1}
\label{tab-rq1-summary-rebuttal}
\begin{tabular}{ p{1.5cm} p{2.5cm} r r r r r }
\hline
\textbf{Lang.} & \textbf{Duration} & \textbf{Repo. Cnt} & \textbf{Dev. Cnt} & \textbf{Com. Cnt} & \textbf{IaC Cnt.} & \textbf{IaC Size}  \\ 
\hline
Ansible & 2014-02 to 2018-11 & 16 & 1,175 & 20,294 & 1,101 & 138,679 \\
Chef    & 2011-05 to 2018-11 & 11 & 650   & 4,758  & 855   & 124,808 \\
\hline
\end{tabular}
\end{table}

%%%%%%%%%%%%%%%%%%%%%%%%%% Rebuttal ::: Summary ::: RQ1 %%%%%%%%%%%%%

\textbf{Methodology Overview}: The security smell derivation process is similar to our prior work~\cite{me:icse2019:slic}, and same for all three languages: Ansible, Chef, and Puppet. First, we collect scripts from an organization who have made their scripts available open source. Next, raters with software security knowledge apply open coding to identify coding patterns that satisfy our definition of security smells. Next, we isolate such coding anti-patterns and assign a category. After assigning a category we check for the CWE database. If a mapping is found then we keep the category, other we discard the category. As we use look for coding patterns and CWE, our security smell categorization process can be applied to any configuration language.

\begin{figure*}[]
\includegraphics[width=0.97\textwidth]{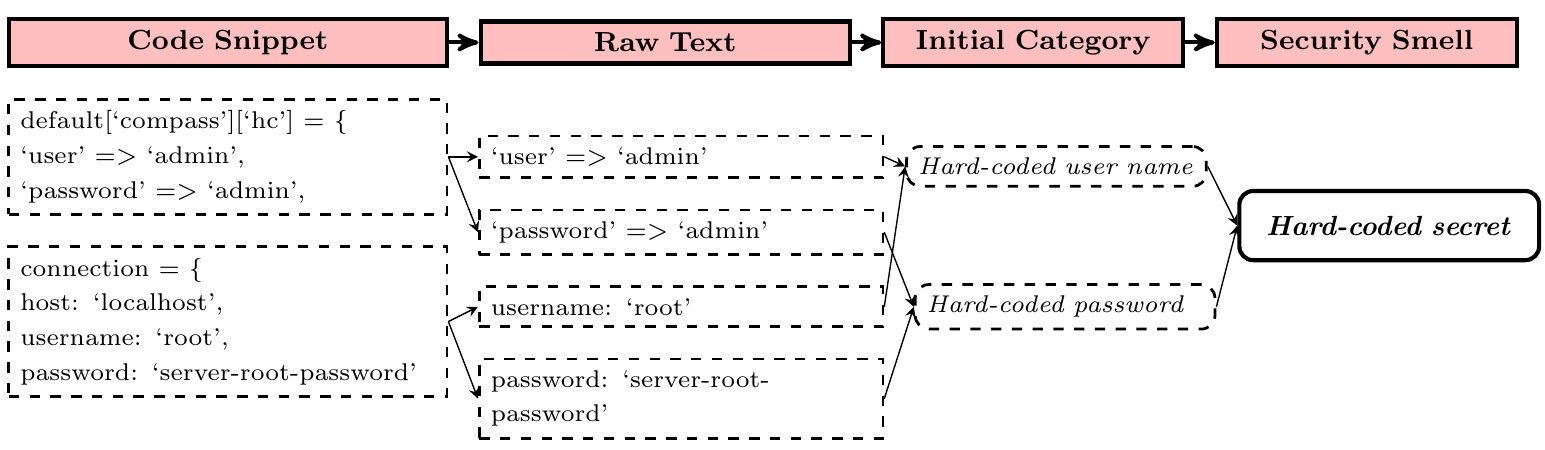}
\caption{An example to demonstrate the process of determining security smells using open coding.}
\label{fig-rq1-cgt}
\end{figure*}

\textbf{Open coding}: We first apply a qualitative analysis technique called open coding~\cite{saldana2015coding} on the collected scripts. In open coding a rater observes and synthesizes patterns within structured or unstructured text~\cite{saldana2015coding}. We select qualitative analysis because we can (i) get a summarized overview of recurring coding patterns that are indicative of security weakness; and (ii) obtain context on how the identified security smells can be automatically identified. We determine security smells by first identifying code snippets that may have security weaknesses based on the first and second author’s security expertise. Figure~\ref{fig-rq1-cgt} provides an example of our qualitative analysis process. We first analyze the code content for each IaC script and extract code snippets that correspond to a security weakness, as shown in Figure~\ref{fig-rq1-cgt}. From the code snippet provided in the top left corner, we extract the raw text: `user' =$>$ `admin'. Next, we generate the initial category `Hard-coded user name' from the raw text ``user' =$>$ `admin'' and `username: `root''. Finally, we determine the smell `Hard-coded secret' by combining initial categories. We combine these two initial categories, as both correspond to a common pattern of specifying user names and passwords as hard-coded secrets. 

Upon derivation of each smell, we map each identified smell to a possible security weakness defined by CWE~\cite{cwe:mitre}.  We select the CWE to map each smell to a security weakness because CWE is a list of common software security weaknesses developed by the security community~\cite{cwe:mitre}. A mapping between a derived security smell and a security weakness reported by CWE can validate our qualitative process. For the example presented in Figure~\ref{fig-rq1-cgt}, we observe the derived security smell `Hard-coded secret' to be related to `CWE-798: Use of Hard-coded Credentials' and `CWE-259: Use of Hard-coded Password'~\cite{cwe:mitre}. Each rater separately mapped each of the identified security smell to an entry in the CWE dictionary.   

%We repeat the above-mentioned process for both 1,101 Ansible and 855 Chef scripts. 

During the time period of conducting open coding, the first author was a PhD student and also the first author of the prior work~\cite{me:icse2019:slic} we replicate. The second author is a PhD student. Both, the first and second author, individually conducted the open coding process. Upon completion of the open coding process, we record the agreements and disagreements for the identified security smells. We also calculate Cohen's Kappa~\cite{cohens:kappa}.          

For Ansible, the first and the second author, respectively, identified four and six security smells. For Chef, the first and the second author respectively identified seven and nine security smells. The Cohen's Kappa is respectively, 0.6 and 0.5 for Ansible and Chef scripts between the first and second author of the paper. The disagreements triggered a discussion session, where both raters’ reasons on why they agreed or disagreed on the identified smell categories. After completing the discussion, both raters individually revisit their categories, and finally both agreed on the set of six and eight security smells respectively, for Ansible and Chef. At this stage the Cohen’s Kappa is 1.0 for both Ansible and Chef. One additional security smell for which both raters agreed upon is `No Integrity Check' for both Ansible and Chef.

\textit{Comments on Generalizability}: Our methodology requires (i) raters with software security experience, (ii) availability of scripts, and (iii) CWE database. As long as these requirements are fulfilled our methodology of deriving smells is generalizable, i.e. can be applied for other IaC languages, such as Terraform. Let us consider a hypothetical example: a researcher wants to replicate our study to derive security smells for Terraform scripts. First step will be using a rater with software security experience. Then, the rater will apply his/her software security knowledge to identify coding patterns and categories. Finally, the rater will check the CWE database if the categories have a direct mapping to the CWE entries.

\subsection{Answer to RQ1: What security smells occur in Ansible and Chef scripts?}
\label{res-rq1}

\begin{figure}[t]
\includegraphics[width=0.95\textwidth]{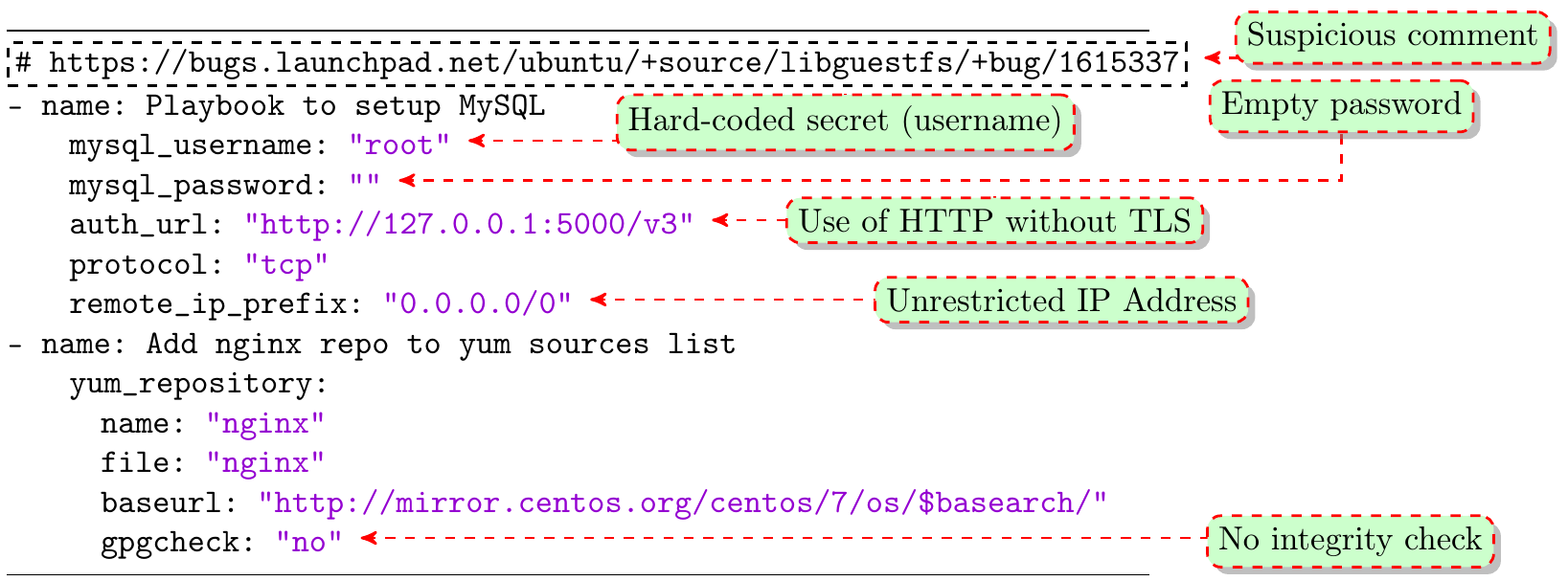}
\caption{An annotated Ansible script with six security smells. The name of each security smell is highlighted on the right.}
\label{fig-rq1-anno-ansi}
\end{figure}

\begin{figure}[]
\includegraphics[width=0.95\textwidth]{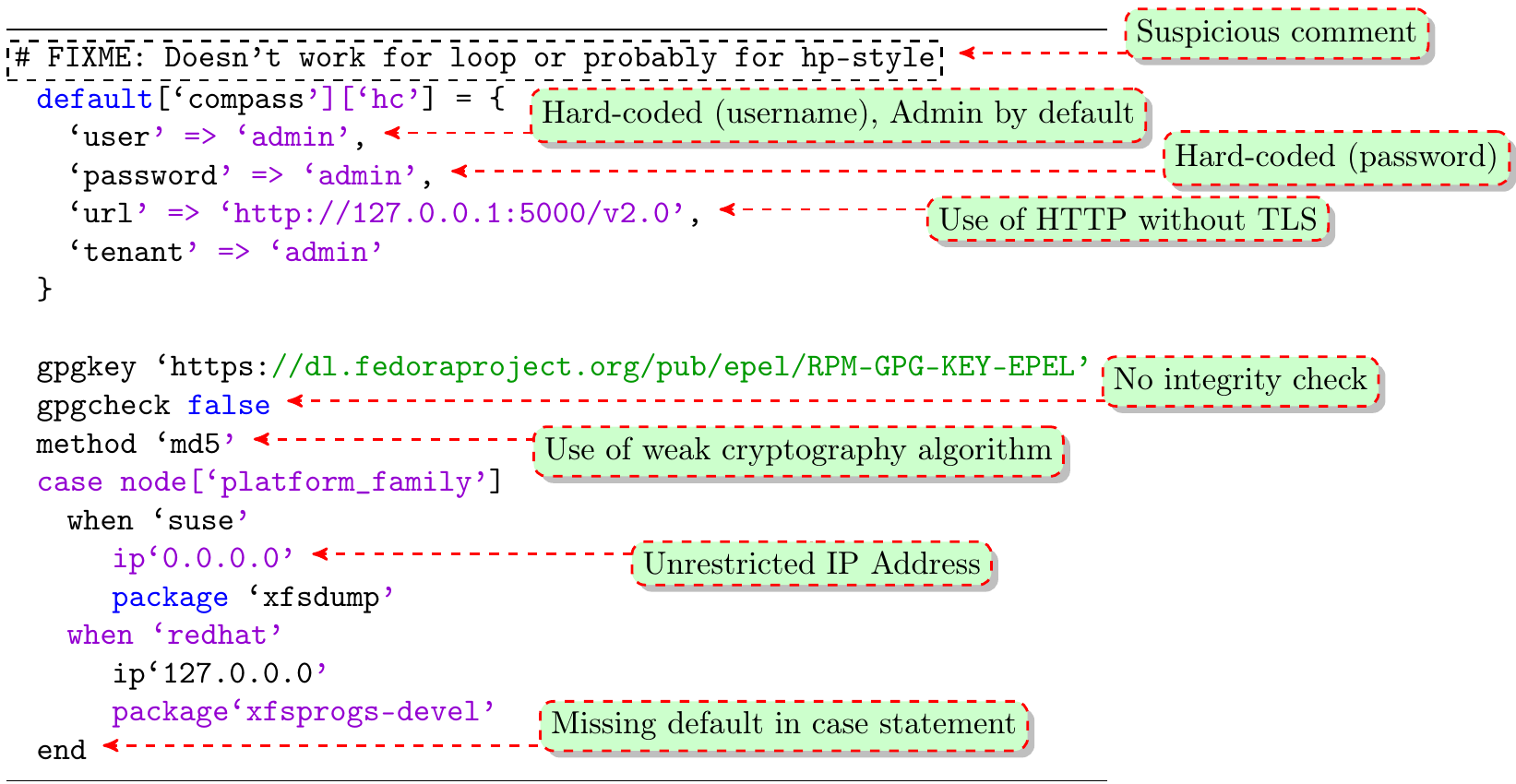}
\caption{An annotated Chef script with eight security smells. The name of each security smell is highlighted on the right.}
\label{fig-rq1-anno-chef}
\end{figure}

We identify six security smells for Ansible scripts: empty password, hard-coded secret, no integrity check, suspicious comment, unrestricted IP address, and use of HTTP Without SSL/TLS. For Chef scripts we identify eight security smells: admin by default, hard-coded secret, no integrity check, suspicious comment, switch statement without default, unrestricted IP address, use of HTTP Without SSL/TLS, and use of weak cryptography algorithm. Rahman et al.~\cite{me:icse2019:slic} identified seven security smells for Puppet scripts: admin by default, empty password, hard-coded secret, suspicious comment, unrestricted IP address, use of HTTP Without SSL/TLS, and use of weak cryptography algorithm. Four security smells are common across all of Ansible, Chef, and Puppet: hard-coded  secret, suspicious comment, unrestricted IP address, and use of HTTP without SSL/TLS. Examples of each security smell for Ansible and Chef are respectively, presented in Figure~\ref{fig-rq1-anno-ansi} and~\ref{fig-rq1-anno-chef}. Below, we list the names of the smells alphabetically, where each smell name is followed by the applicable language: Ansible (\includegraphics[width=0.35cm, height=0.35cm]{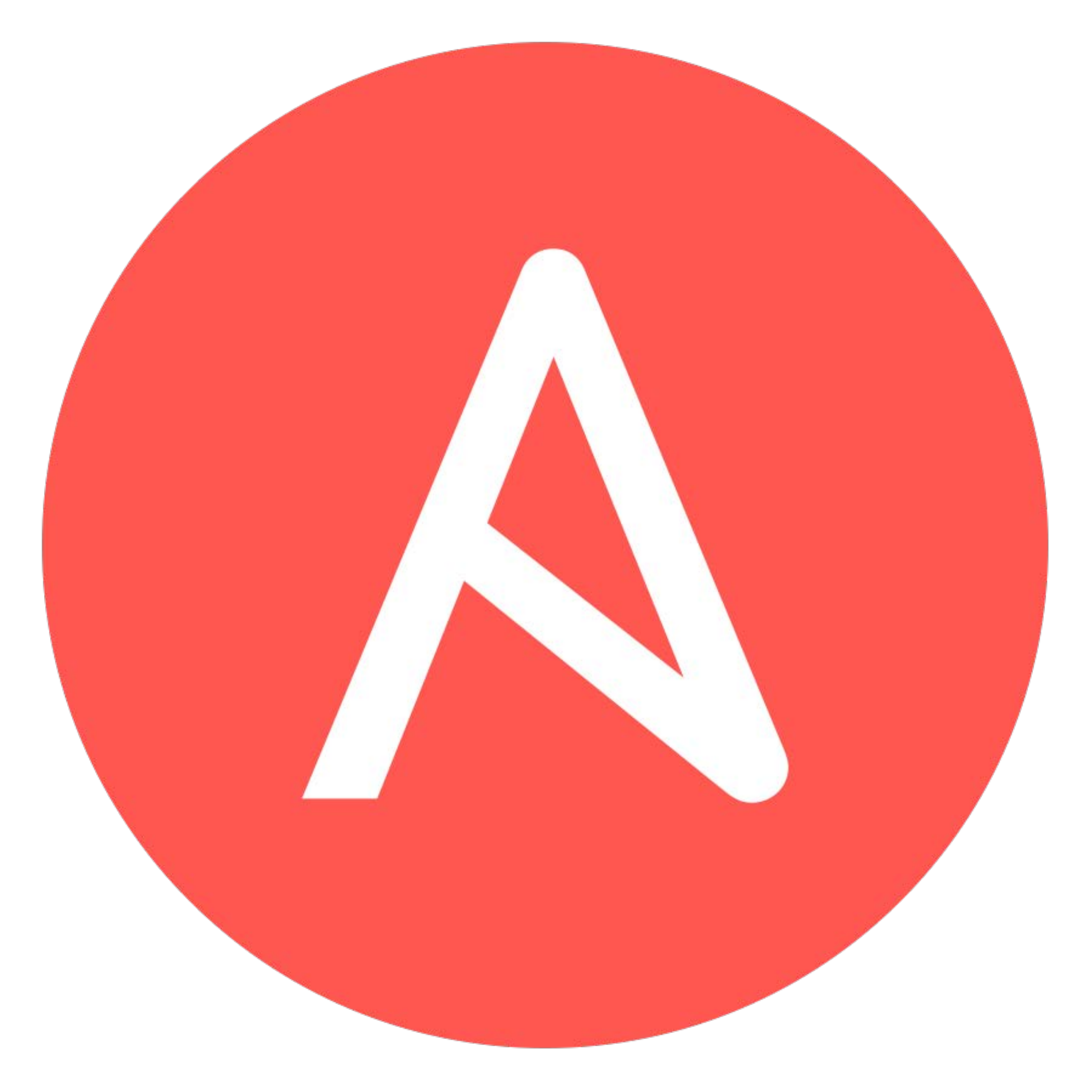}) and Chef (\includegraphics[width=0.35cm, height=0.35cm]{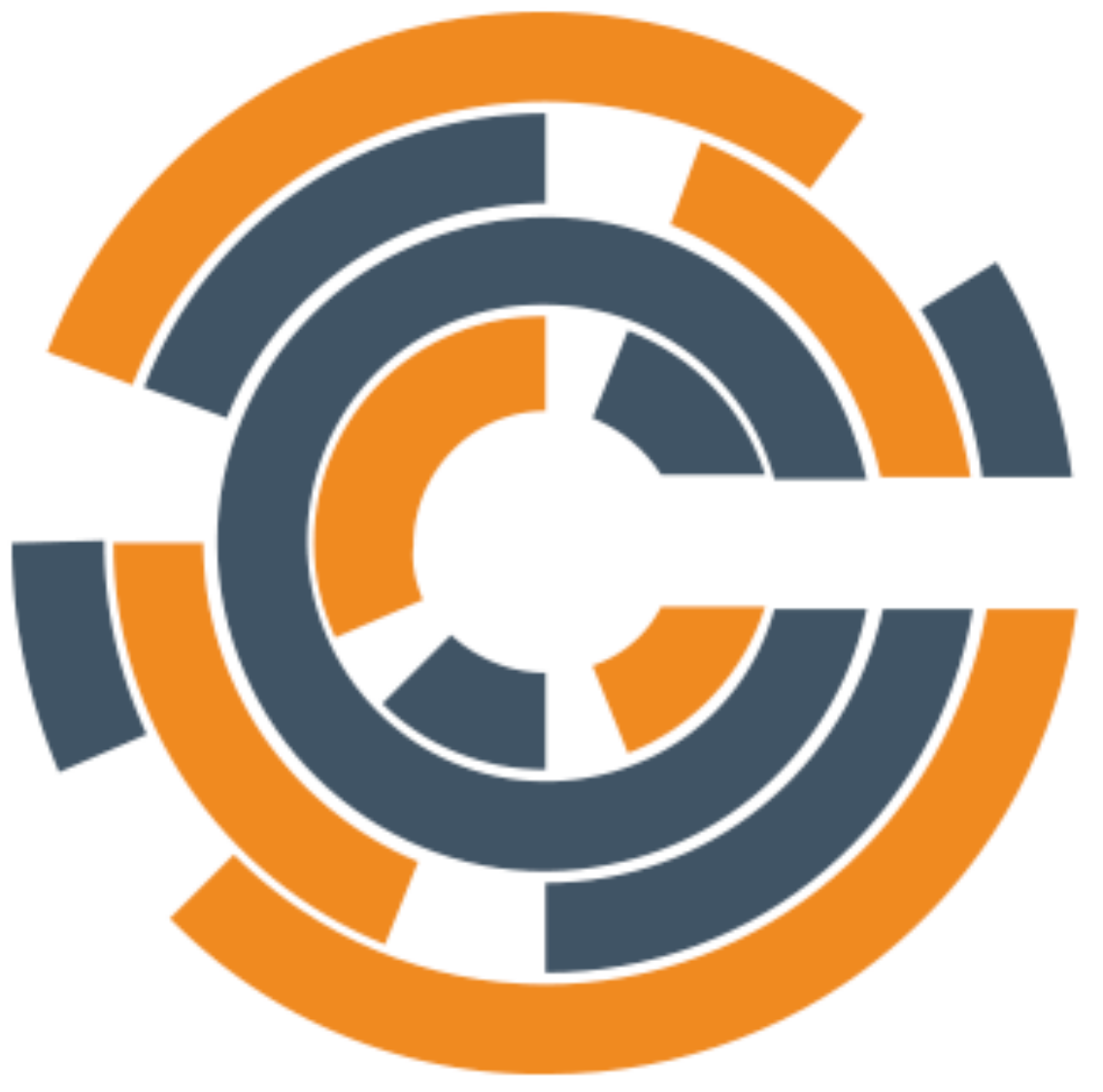}). 

\textbf{Admin by default} (\includegraphics[width=0.35cm, height=0.35cm]{chef.pdf}) :\label{rq1-adm} This smell is the recurring pattern of specifying default users as administrative users. The smell can violate the `principle of least privilege' property~\cite{nist2014_least_privi}, which recommends practitioners design and implement a system in a manner so that by default the least amount of access necessary is provided to any entity. In Figure~\ref{fig-rq1-anno-chef}, an `admin' user will be created in the `default' mode of provisioning an infrastructure. The smell is related with `CWE-250: Execution with Unnecessary Privileges'~\cite{cwe:mitre}. 

\textbf{Empty password} (\includegraphics[width=0.35cm, height=0.35cm]{ansible.pdf}) :\label{rq1-empt-pass} This smell is the recurring pattern of using a string of length zero for a password. An empty password is indicative of a weak password. An empty password does not always lead to a security breach, but makes it easier to guess the password. In SSH key-based authentication, instead of passwords, public and private keys can be used~\cite{ssh:orig}. Our definition of empty password does not include usage of no passwords and focuses on attributes/variables that are related to passwords and assigned an empty string. Empty passwords are not included in hard-coded secrets because for a hard-coded secret, a configuration value must be a string of length one or more. The smell is similar to the weakness `CWE-258: Empty Password in Configuration File'~\cite{cwe:mitre}.       

%An empty password is different from using no passwords

\textbf{Hard-coded secret} (\includegraphics[width=0.35cm, height=0.35cm]{ansible.pdf} \includegraphics[width=0.35cm, height=0.35cm]{chef.pdf}) :\label{rq1-hard} This smell is the recurring pattern of revealing sensitive information, such as user name and passwords in IaC scripts. IaC scripts provide the opportunity to specify configurations for the entire system, such as configuring user name and password, setting up SSH keys for users, specifying authentications files (creating key-pair files for Amazon Web Services). However, programmers can hard-code these pieces of information into scripts. We consider three types of hard-coded secrets: hard-coded passwords, hard-coded user names, and hard-coded private cryptography keys. We acknowledge that practitioners may intentionally leave hard-coded secrets, such as user names and SSH keys in scripts, which may not be enough to cause a security breach. Hence this practice is security smell, but not a vulnerability. Relevant weaknesses to the smell are `CWE-798: Use of Hard-coded Credentials' and `CWE-259: Use of Hard-coded Password'~\cite{cwe:mitre}.    

\textbf{Missing Default in Case Statement}  (\includegraphics[width=0.35cm, height=0.35cm]{chef.pdf}) :\label{rq1-case-default} This smell is the recurring pattern of not handling all input combinations when implementing a case conditional logic. Because of this coding pattern, an attacker can guess a value, which is not handled by the case conditional statements and trigger an error. Such an error can provide the attacker unauthorized information for the system in terms of stack traces or system error. This smell is related to `CWE-478: Missing Default Case in Switch Statement'~\cite{cwe:mitre}.   

\textbf{No integrity check} (\includegraphics[width=0.35cm, height=0.35cm]{ansible.pdf} \includegraphics[width=0.35cm, height=0.35cm]{chef.pdf}) :\label{rq1-integ-check} This smell is the recurring pattern of downloading content from the Internet and not checking the downloaded content using checksums or gpg signatures. We observe the following type of content downloaded from the Internet without checking for integrity: .tar, .tgz, .tar.gz, .dmg, .rpm, and .zip. By not checking for integrity, a developer assumes the downloaded content is secure and has not been corrupted by a potential attacker. Checking for integrity provides an additional layer of security to ensure that the downloaded content is intact, and the downloaded link has not been compromised by an attacker, possibly inserting a virus payload. This smell is related to `CWE-353: Missing Support for Integrity Check'~\cite{cwe:mitre}. 

\textbf{Suspicious comment} (\includegraphics[width=0.35cm, height=0.35cm]{ansible.pdf} \includegraphics[width=0.35cm, height=0.35cm]{chef.pdf}) :\label{rq1-susp} This smell is the recurring pattern of putting information in comments about the presence of defects, missing functionality, or weakness of the system. Examples of such comments include putting keywords such as `TODO', `FIXME', and `HACK' in comments, along with putting bug information in comments. Keywords such as `TODO' and `FIXME' in comments are used to specify an edge case or a problem~\cite{ICSE:TODO:Storey}. However, these keywords make a comment `suspicious'. The smell is related to `CWE-546: Suspicious Comment'~\cite{cwe:mitre}. 

\textbf{Unrestricted IP Address} (\includegraphics[width=0.35cm, height=0.35cm]{ansible.pdf} \includegraphics[width=0.35cm, height=0.35cm]{chef.pdf}) :\label{rq1-ip} This smell is the recurring pattern of assigning the address 0.0.0.0 for a database server or a cloud service/instance. Binding to the address 0.0.0.0 may cause security concerns as this address can allow connections from every possible network~\cite{attack:bind:dos}. Such binding can cause security problems as the server, service, or instance will be exposed to all IP addresses for connection. For example, practitioners have reported how binding to 0.0.0.0 facilitated security problems for MySQL~\footnote{https://serversforhackers.com/c/mysql-network-security}(database server), Memcached~\footnote{https://news.ycombinator.com/item?id=16493480}(cloud-based cache service) and Kibana~\footnote{https://www.elastic.co/guide/en/kibana/5.0/breaking-changes-5.0.html}(cloud-based visualization service). We acknowledge that an organization can opt to bind a database server or cloud instance to 0.0.0.0, but this case may not be desirable overall. This security smell has been referred to as `Invalid IP Address Binding' in our prior work~\cite{me:icse2019:slic}. This smell is related to improper access control as stated in the weakness `CWE-284: Improper Access Control'~\cite{cwe:mitre}.

\textbf{Use of HTTP without SSL/TLS}  (\includegraphics[width=0.35cm, height=0.35cm]{ansible.pdf} \includegraphics[width=0.35cm, height=0.35cm]{chef.pdf}) :\label{rq1-http} This smell is the recurring pattern of using HTTP without the Transport Layer Security (TLS) or Secure Sockets Layer (SSL). Such use makes the communication between two entities less secure, as without SSL/TLS, use of HTTP is susceptible to man-in-the-middle attacks~\cite{attack:http:mitm}. For example, as shown in Figure~\ref{fig-rq1-anno-ansi}, the authentication URL uses HTTP without SSL/TLS for `auth\_url'. Such usage of HTTP can be problematic, as an attacker can eavesdrop on the communication channel. Information sent over HTTP may be encrypted, and in such case `Use of HTTP without SSL/TLS' may not lead to a security attack. We have referred to this security smell as `Use of HTTP without TLS' in our prior work~\cite{me:icse2019:slic}. This security smell is related to `CWE-319: Cleartext Transmission of Sensitive Information'~\cite{cwe:mitre}.  

\textbf{Use of weak cryptography algorithms} (\includegraphics[width=0.35cm, height=0.35cm]{chef.pdf}) :\label{rq1-md5} This smell is the recurring pattern of using weak cryptography algorithms, namely, MD5 and SHA-1, for encryption purposes. MD5 suffers from security problems, as demonstrated by the Flame malware in 2012~\cite{flame:md5}. MD5 is susceptible to collision attacks~\cite{attack:md5:collision} and modular differential attacks~\cite{attack:md5:md}. Similar to MD5, SHA1 is also susceptible to collision attacks~\footnote{https://security.googleblog.com/2017/02/announcing-first-sha1-collision.html}. Using weak cryptography algorithms for hashing that may not always lead to a breach. However, using weak cryptography algorithms for setting up passwords may lead to a breach. This smell is related to `CWE-327: Use of a Broken or Risky Cryptographic Algorithm' and `CWE-326: Inadequate Encryption Strength'~\cite{cwe:mitre}.  

%%%%%%%%%%%%%%%%%%%%%%%%%%%%%%%%%%%%%%%%%%%%%%%%% SLAC %%%%%%%%%%%%%%%%%%%%%%%%%%%%%%%%%%%%%%%%%%%%%%%%%%%%%%%%
%%%%%%%%%%%%%%%%%%%%%%%%%%%%%%%%%%%%%%%%%%%%%%%%%%%%%%%%%%%%%%%%%%%%%%%%%%%%%%%%%%%%%%%%%%%%%%%%%%%%%%%%%%%%%%%

\section{Security Linter for Ansible and Chef Scripts (SLAC)}
\label{slic} 

We construct Security Linter for Ansible and Chef Scripts (SLAC) to help practitioners automatically identify security smells in Ansible and Chef scripts. We first describe how we constructed SLAC, then we describe how we evaluated SLAC's smell detection accuracy. 

\subsection{Description of SLAC}
\label{meth-slic-desc}

\begin{figure}[htb]
\subfloat[]{
 \includegraphics[width=0.45\textwidth]{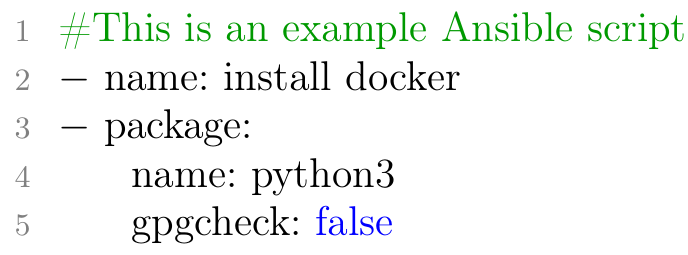}
 \label{fig-meth-parser-exa-script-ansi}
}
\subfloat[]{
 \includegraphics[width=0.49\textwidth]{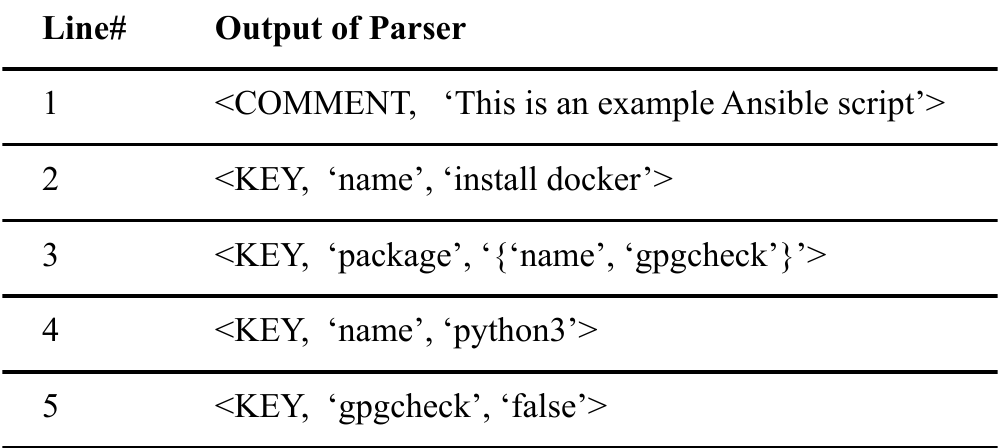}
 \label{fig-meth-parser-exa-parser-ansi}
}
\caption{Output of the `Parser' component in SLAC. Figure~\ref{fig-meth-parser-exa-script-ansi} presents an example Ansible script fed to Parser. Figure~\ref{fig-meth-parser-exa-parser-ansi} presents the output of Parser for the example Ansible script.} 
\label{fig-meth-parser-ansi} 
\end{figure}

\begin{figure}[htb]
\subfloat[]{
 \includegraphics[width=0.45\textwidth]{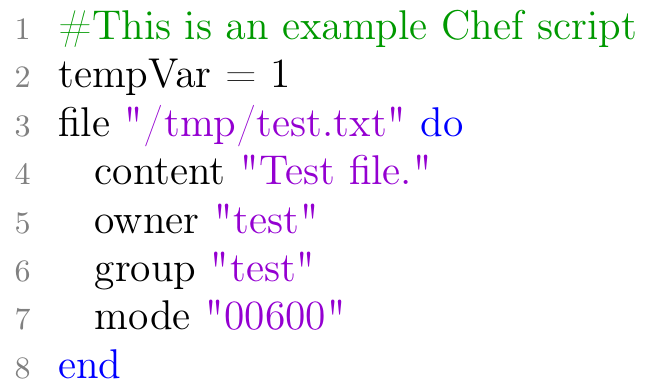}
 \label{fig-meth-parser-exa-script-chef}
}
\subfloat[]{
 \includegraphics[width=0.49\textwidth]{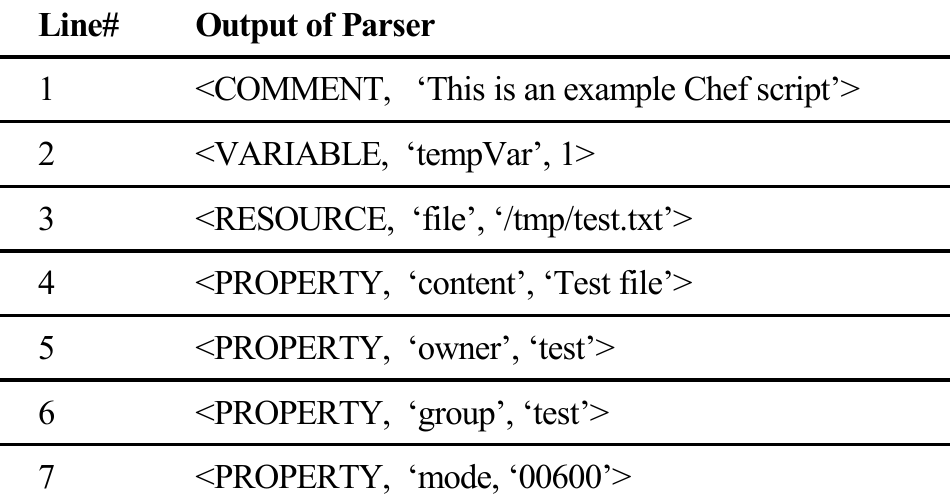}
 \label{fig-meth-parser-exa-parser-chef}
}
\caption{Output of the `Parser' component in SLAC. Figure~\ref{fig-meth-parser-exa-script-chef} presents an example Chef script fed to Parser. Figure~\ref{fig-meth-parser-exa-parser-chef} presents the output of Parser for the example Chef script.} 
\label{fig-meth-parser-chef} 
\end{figure}

SLAC is a static analysis tool for detecting the six and eight security smells respectively, for Ansible and Chef scripts. SLAC has two extensible components: 

\textbf{Parser}: The Parser parses an Ansible or Chef script and returns a set of tokens. Tokens are non-whitespace character sequences extracted from IaC scripts, such as keywords and variables. Except for comments, each token is marked with its name, token type, and any associated configuration value. Only token type and configuration value are marked for comments. For example, Figures~\ref{fig-meth-parser-exa-script-ansi} and~\ref{fig-meth-parser-exa-script-chef} respectively provides a sample script in Ansible and Chef that is fed into SLAC. The output of Parser is expressed as a vector, as shown in Figures~\ref{fig-meth-parser-exa-parser-ansi} and~\ref{fig-meth-parser-exa-parser-chef}. For example in Figure~\ref{fig-meth-parser-exa-parser-chef}, the comment in line\#1, is expressed as the vector `\textless COMMENT, `This is an example Chef script'\textgreater'. 

In the case of Ansible, Parser first identifies comments. Next, for non-commented lines Parser uses a YAML parser and constructs a nested list of key-values pairs in JSON format. We use these key-value pairs to construct rules for the Rule Engine.  

%The Parser parses an IaC script and returns a set of tokens. Tokens are non-whitespace character sequences extracted from IaC scripts, such as keywords and variables. 

Similar to Ansible, in the case of Chef, Parser first identifies comments. Next, Parser identifies each token in a Chef script is marked with its name, token type, and any associated configuration value. For example, Figure~\ref{fig-meth-parser-exa-script-chef} provides a sample script that is fed into SLAC.  The output of Parser is is expressed as a vector, as shown in Figure~\ref{fig-meth-parser-exa-parser-chef}. For example, the comment in line\#1, is expressed as the vector `\textless COMMENT, `This is an example Chef script'\textgreater'. Parser provides a vector representation of all code snippets in a script.

\textbf{Rule Engine}: Following the study design of prior work~\cite{me:icse2019:slic}, we use a rule-based approach to detect security smells. We use rules because (i) unlike keyword-based searching, rules are less susceptible to false positives~\cite{icomment:sosp07}; and (ii) rules can be applicable for IaC tools irrespective of their syntax. The Rule Engine consists of a set of rules that correspond to the set of security smells identified in Section~\ref{meth-cgt}. The Rule Engine uses the set of tokens extracted by Parser and checks if any rules are satisfied. 

\begin{table*}[htb]
\centering
\caption{An Example of Using Code Snippets To Determine Rule for `Use of HTTP Without SSL/TLS'}
\label{table-meth-rule-exa}
\begin{tabular}{  p{6.5cm}  p{7.5cm}  }
\hline
\textbf{Code Snippets} & \textbf{Output of Parser}   \\
\hline
repo=`http://ppa.launchpad.net/chris-lea/node.js-legacy/ubuntu' & \textless VARIABLE, `repo', `http://ppa.launchpad.net/chris-lea/node.js-legacy/ubuntu' \textgreater \\
repo=`http://ppa.launchpad.net/chris-lea/node.js/ubuntu' & \textless VARIABLE, `repo', `http://ppa.launchpad.net/chris-lea/node.js/ubuntu' \textgreater  \\
auth\_uri=`http://localhost:5000/v2.0' & \textless VARIABLE, `auth\_uri', `http://localhost:5000/v2.0' \textgreater   \\
uri `http://binaries.erlang-solutions.com/debian' & \textless PROPERTY, `uri', `http://binaries.erlang-solutions.com/debian' \textgreater  \\
url `http://pkg.cloudflare.com' & \textless PROPERTY, `url', `http://pkg.cloudflare.com' \textgreater  \\
\hline
\end{tabular}
\end{table*}

We can identify properties of source code from the smell-related code snippets and constitute rules using the source code properties. Each smell-related code snippet can show what properties of a script is related with a security smell occurrence. We use Table~\ref{table-meth-rule-exa} to demonstrate our approach. The `Code Snippet' column presents a list of code snippets related to `Use of HTTP without SSL/TLS'. The `Parser Output' column represents vectors for each code snippet. We observe that the vector of format `\textless VARIABLE, NAME, CONFIGURATION VALUE \textgreater' and `\textless PROPERTY, NAME, CONFIGURATION VALUE \textgreater', respectively, occurs three times and twice for our example set of code snippets. We use the vectors from the output of `Parser' to determine that variable and properties are related to `Use of HTTP without SSL/TLS'. The vectors can be abstracted to construct the following rule: `($isVariable(x)$ $\vee$ $isProperty(x)$) $\wedge$ $isHTTP(x)$'. This rule states that `for an IaC script, if token $x$ is a variable or a property, and a string is passed as configuration value for a variable or a property which is related to specifying a URL that uses HTTP without SSL/TLS support, then the script contains the security smell `Use of HTTP without SSL/TLS'. We apply the process of abstracting patterns from smell-related code snippets to determine the rules for the all security smells for both Ansible and Chef.       

A programmer can use SLAC to identify security smells for one or multiple Ansible and Chef scripts. The programmer specifies a directory where script(s) reside. Upon completion of analysis, SLAC generates a comma separated value (CSV) file where the count of security smell for each script is reported. We implement SLAC using API methods provided by PyYAML~\footnote{https://pyyaml.org/} for Ansible and  Foocritic~\footnote{http://www.foodcritic.io/} for Chef.  

\textbf{Rules to Detect Security Smells}: For Ansible and Chef we present the rules needed for the `Rule Engine' of SLAC respectively in Tables~\ref{tab-res-ansi-rules} and~\ref{tab-res-chef-rules}. The string patterns needed to support the rules in Tables~\ref{tab-res-ansi-rules} and~\ref{tab-res-chef-rules} are listed in Table~\ref{tab-res-slic-kw}. The `Rule' column lists rules for each smell that is executed by Rule Engine to detect smell occurrences. To detect whether or not a token type is a resource ($isResource(x)$), a property ($isProperty(x)$), or a comment ($isComment(x)$), we use the token vectors generated by Parser. Each rule includes functions whose execution is dependent on matching of string patterns. We apply a string pattern-based matching strategy similar to prior work~\cite{Bosu:VCC2014}~\cite{Bugiel:AmazonIA:CCS2015}, where we check if the value satisfies the necessary condition. Table~\ref{tab-res-slic-kw} lists the functions and corresponding string patterns. For example, function `hasBugInfo()' will return true if the string pattern `show\_bug\textbackslash.cgi?id=[0-9]+' or `bug[\#$\textbackslash t$]$\ast$[0-9]+' is satisfied. 

For Ansible and Chef scripts the rule engine takes output from the Parser, and checks if any of the rules listed in Tables~\ref{tab-res-ansi-rules} and~\ref{tab-res-chef-rules} respectively, for Ansible and Chef. In Tables~\ref{tab-res-ansi-rules} and~\ref{tab-res-chef-rules} the `Rule' column lists rules for each smell that is executed by Rule Engine to detect smell occurrences. In the case of Ansible scripts, we used the output from Parser to obtain the key value pairs ($k$, $k.value$) and comments needed to execute the rules listed in Table~\ref{tab-res-ansi-rules}. Similarly, in the case of Chef scripts, we use the output of Parser to check variables ($isVariable(x)$), properties ($isProperty(x)$), attributes ($isAttribute(x)$), and case statements ($isCaseStmt(x)$). Each rule includes functions whose execution is dependent on matching of string patterns.

%%%%%%%%%%%%%%%%%%%Ansible Rule Table %%%%%%%%%%%%%%%%%%%%%%%%%%%%%%%%%%
\begin{table}[]
\centering
\caption{Rules to Detect Security Smells for Ansible Scripts} 
\label{tab-res-ansi-rules}
\begin{tabular}{ p{3.5cm}  p{7.5cm} }
\hline
\multicolumn{1}{c}{\textbf{Smell Name}} & \multicolumn{1}{c}{\textbf{Rule}}                                                                                                                                                   \\
\hline 
Empty password                          & ( $isKey(k)$ $\wedge$ $length(k.value)==0$ $\wedge$ $isPassword(k)$ ) \\
\hline 
Hard-coded secret                       & \shortstack[l]{($isKey(k)$ $\wedge$ $length(k.value)\textgreater 0$) $\wedge$ \\ ($isUser(k)$ $\vee$ $isPassword(k)$ $\vee$ $isPrivateKey(k)$)}                                     \\
\hline 
No integrity check                      & \shortstack{ ($isKey(k)$ $\wedge$ \\ ($isIntegrityCheck(x)==False$ $\wedge$ $isDownload(x.value)$) ) }\\
\hline 
Suspicious comment                      & ( $isComment(k)$ $\wedge$ $hasWrongWord(k)$ $\vee$ $hasBugInfo(k)$ ) \\
\hline 
Unrestricted IP address                 & ($isKey(k)$ $\wedge$ $isInvalidBind(k.value)$) \\
\hline 
Use of HTTP without SSL/TLS             & ( $isKey(k)$ $\wedge$ $isHTTP(k.value)$ )   \\
\hline                                                                                                                       
\end{tabular}
\end{table}

%%%%%%%%%%%%%%%%%%%Chef Rule Table %%%%%%%%%%%%%%%%%%%%%%%%%%%%%%%%%%

\begin{table*}[]
\centering
\caption{Rules to Detect Security Smells for Chef Scripts}
\label{tab-res-chef-rules}
\begin{tabular}{  p{3.5cm}  p{10.5cm}  }
\hline
\textbf{Smell Name} & \textbf{Rule}   \\
\hline
Admin by default & \shortstack{($isPropertyOfDefaultAttribute(x)$) $\wedge$ \\ (($isAdmin(x.name)$) $\wedge$ ($isUser(x.name)$ $\vee$ $isRole(x.name)$ ))}\\
\hline
Hard-coded secret & \shortstack[l]{($isProperty(x)$ $\vee$ $isVariable(x)$) $\wedge$ ($isUser(x.name)$ \\ $\vee$ $isPassword(x.name)$ $\vee$ $isPvtKey(x.name)$) \\ $\wedge$ ($length(x.value)$\textgreater$0$)} \\
\hline
Missing default in case & \shortstack{($isCaseStmt(x)$ $\wedge$ $x.elseBranch==False$)}\\
\hline 
No integrity check & \shortstack{ ($isProperty(x)$ $\vee$ $isAttribute(x)$) $\wedge$ \\ ($isIntegrityCheck(x)==False$ $\wedge$ $isDownload(x.value)$)  }\\
\hline
Suspicious comment & \shortstack{($isComment(x)$) $\wedge$ ($hasWrongWord(x)$ $\vee$ $hasBugInfo(x)$)}\\
\hline
Unrestricted IP address & \shortstack{(($isVariable(x)$ $\vee$ $isProperty(x)$) $\wedge$ ($isInvalidBind(x.value)$)}\\
\hline
Use of HTTP without SSL/TLS &   (\shortstack{$isProperty(x)$ $\vee$ $isVariable(x)$) $\wedge$ ($isHTTP(x.value)$} ) \\
\hline
Use of weak crypto. algo. & (\shortstack{$isAttribute(x)$ $\wedge$ $usesWeakAlgo(x.value)$} )\\
\hline
\end{tabular}
\end{table*}

\begin{table}[htb]
\centering
\caption{String Patterns Used for Functions in Rules}
\label{tab-res-slic-kw}
\begin{tabular}{  p{3cm}  p{9.5cm}  }
\hline
\textbf{Function} & \textbf{String Pattern}   \\
\hline
$hasBugInfo()$~\cite{zeller:bug:regex} & `bug[\#$\textbackslash t$]$\ast$[0-9]+',`show\_bug\textbackslash.cgi?id=[0-9]+'\\
\hline
$hasWrongWord()$~\cite{cwe:mitre} & `bug', `hack', `fixme', `later', `later2', `todo'\\
\hline
$isAdmin()$ & `admin'\\
\hline
$isDownload()$ & `http[s]?://(?:[a-zA-Z]$|$[0-9]$|$[\$-\_@.\&+]$|$[!*\(\),]$|$(?:\%[0-9a-fA-F][0-9a-fA-F]))+.[dmg$|$rpm$|$tar.gz$|$tgz$|$zip$|$tar]' \\
\hline
$isHTTP()$ &  `http:' \\
\hline
$isInvalidBind()$ &  `0.0.0.0' \\
\hline
$isIntegrityCheck()$ &  `gpgcheck', `check\_sha', `checksum', `checksha' \\
\hline
$isPassword()$ & `pwd', `pass', `password' \\
\hline
$isPvtKey()$ & `[pvt$\arrowvert$priv]+*[cert$\arrowvert$key$\arrowvert$rsa$\arrowvert$secret$\arrowvert$ssl]+'\\
\hline
$isRole()$ & `role'\\
\hline
$isUser()$ & `user'\\
\hline
$usesWeakAlgo()$ &  `md5', `sha1' \\
\hline
\end{tabular}
\end{table}

\subsection{Evaluation of SLAC}
\label{meth-slic-eval}

We use raters to construct the oracle dataset to mitigate author bias in SLAC's evaluation, similar to Chen et al.~\cite{chen:icse2017:logging} and our prior work~\cite{me:icse2019:slic}. We construct four oracle datasets in two rounds. In the first round we use graduate students from NC State University in March and April 2019. In the second round we sue a third year PhD student from Tennessee Technological University to construct oracle datasets for Ansible and Chef. In the second round, we don not include Ansible and Chef scripts that are included and analyzed in the first round. We describe oracle dataset construction process for both round in the following subsections:

\subsubsection{Evaluation of SLAC in Round\#1}
\label{slic-eval-round1}

We first provide the process of oracle data construction. Next, we provide the performance of SLAC. 

\textbf{Round\#1-Oracle Dataset for Ansible and Chef}: For each of Ansible and Chef, we construct an oracle datasets using closed coding~\cite{saldana2015coding}, where at least two raters identifies a pre-determined pattern, and their agreement is checked. 

We used graduate students as raters to construct the oracle dataset. We recruited these raters from a graduate-level course related to DevOps conducted in March and April of 2019 at NC State University. Of the 60 students in the class, 32 students agreed to participate. The raters apply their knowledge related to IaC scripts and software security to determine if a certain smell appears for a script.  We assigned 96 Ansible and 76 Chef scripts scripts to the 32 students to ensure each script is reviewed by at least two students.  The scripts were selected randomly from the 16 Ansible and 10 Chef repositories, respectively, for Ansible and Chef. Each student did not have to rate more than 15 scripts. Prior to allocating the assignments to the students, we obtained Institutional Review Board (IRB) approval (IRB\# 12563).  

We made the smell identification task available to the raters using a website~\footnote{http://13.59.115.46/website/start.php}. The website includes a handbook on Ansible and Chef, and a document that shows examples of security smell instances for both Ansible and Chef. In each task, a rater determines which of the six and eight security smells identified in Section~\ref{meth-cgt} occur, respectively, for Ansible and Chef scripts.   The graduate students may miss instances of security smells. To mitigate this limitation, after the students conducted closed coding, the first author conducted manual analysis of the 96 Ansible and 76 Chef scripts to identify if security smells have been missed by the raters.      

We used balanced block design, a technique to randomly allocate items between multiple categories~\cite{balanced:block:design:original}, to assign 96 Ansible and 76 Chef scripts. For Ansible, we observe agreements on the rating for 64 of 96 scripts (66.7\%), with a Cohen's Kappa of 0.4. For Chef, we observe agreements on the rating for 61 of 76 scripts (80.2\%), with a Cohen's Kappa of 0.5. According to Landis and Koch's interpretation~\cite{Landis:Koch:Kappa:Range}, the reported agreement is `fair' and `moderate' respectively, for Ansible and Chef. 

After quantifying the agreement rate, the first author manually inspected 64 Ansible and 61 Chef scripts, for which students agreed. During the manual inspection process, the first author did not use SLAC to identify security smell occurrences. The first author found 17 and 41 security smell occurrences missed by the students, respectively, for Ansible and Chef. The first author added the 17 Ansible and 41 Chef security smells occurrences to the oracle dataset. 

Next, the first author resolved disagreements for 32 Ansible scripts and 15 Chef scripts. The disagreements amongst raters occurred for two reasons: (i) students disagreed on the category, and (ii) students disagreed on presence of security smells. After resolving disagreements, and inspecting scripts for which students agreed upon, we obtain an oracle of 24 Ansible and 67 Chef smell occurrences, as listed in the `Occu.' column of Tables~\ref{tab-res-slic-ansi-accu} and~\ref{tab-res-slic-chef-accu}. Of the 24 Ansible and 67 Chef smell occurrences, respectively, 7 and 26 smells were identified by the students.  

Upon completion of the oracle dataset, we run SLAC for the oracle dataset. Next, we evaluate the accuracy of SLAC using precision and recall for the oracle dataset. Precision refers to the fraction of correctly identified smells among the total identified security smells, as determined by SLAC. Recall refers to the fraction of correctly identified smells that have been retrieved by SLAC over the total amount of security smells. 

\textbf{Round\#1-Performance of SLAC for Ansible and Chef Oracle Dataset}: We report the detection accuracy of SLAC with respect to precision and recall for Ansible in Table~\ref{tab-res-slic-ansi-accu} and Chef in Table~\ref{tab-res-slic-chef-accu}. As shown in the `No smell' row, we identify 77 Ansible scripts with no security smells. The detection accuracy in Tables~\ref{tab-res-slic-ansi-accu} and~\ref{tab-res-slic-chef-accu} corresponds to the accuracy of detecting security smell instances. Along with reporting SLAC's detection accuracy for the oracle dataset, we also report SLAC's detection accuracy for the 7 and 26 security smells identified by the students respectively, in Tables~\ref{tab-res-slac-ansi-student} and~\ref{tab-res-slac-chef-student}. Tables~\ref{tab-res-slic-chef-accu} and~\ref{tab-res-slac-ansi-student} summarizes accuracy respectively, for the complete Chef oracle dataset, and the Ansible security smells identified by the students. For Tables~\ref{tab-res-slac-ansi-student} and~\ref{tab-res-slac-chef-student}, students disagreed upon security smell occurrences and categories. The disagreements were resolved by the first author.

\begin{table}[htb]
\centering
\caption{SLAC's Accuracy for the Ansible Oracle Dataset (Round\#1)}
\label{tab-res-slic-ansi-accu}
\begin{tabular}{ p{4.5cm} r r r }
\hline
\textbf{Smell Name} & \textbf{Occurr.} & \textbf{Precision} & \textbf{Recall} \\ 
\hline
Empty password              & 1  & 1.0 & 1.0 \\ 
Hard-coded secret           & 1  & 1.0 & 1.0 \\
No Integrity Check          & 2  & 1.0 & 1.0 \\ 
Suspicious comment          & 4  & 1.0 & 1.0 \\ 
Unrestricted IP address     & 2  & 1.0 & 1.0 \\ 
Use of HTTP without SSL/TLS & 14 & 1.0 & 1.0 \\
\hline
No smell                    & 77 & 1.0 & 1.0 \\ 
\hline
Average &                       & 1.0 & 1.0 \\
\hline
\end{tabular}
\end{table}

\begin{table}[htb]
\centering
\caption{SLAC's Accuracy for the Chef Oracle Dataset (Round\#1)}
\label{tab-res-slic-chef-accu}
\begin{tabular}{  p{4.5cm}  r  r  r  }
\hline
\textbf{Smell Name} & \textbf{Occurr.}  & \textbf{Precision} & \textbf{Recall}  \\
\hline
Admin by default            & 2  &  1.0  &  1.0 \\
Hard-coded secret           & 25 &  0.8  &  1.0 \\
Suspicious comment          & 10 &  1.0  &  1.0  \\
Unrestricted IP address     & 1  &  1.0  &  1.0 \\
Use of HTTP without SSL/TLS & 27 &  1.0  &  1.0 \\
Use of weak crypto. algo.   & 2  &  1.0  &  1.0 \\
\hline
No smell                    & 61 &  1.0  &  0.9 \\
\hline
Average                     &    & 0.9   &  0.9  \\
\hline
\end{tabular}
\end{table}

%%%%%%%%%%%%%%%%%%%%%%%%%%% Student oracle table Ansible start %%%%%%%%%%%%%%%%%%%%

\begin{table}[htb]
\centering
\caption{SLAC's Accuracy for Ansible Security Smell Occurrences Identified Only by Students}
\label{tab-res-slac-ansi-student}
\begin{tabular}{ p{4cm} r r r }
\hline
\textbf{Smell Name} & \textbf{Occurr.} & \textbf{Precision} & \textbf{Recall} \\ 
\hline
Suspicious comment          & 1  & 1.0 & 1.0 \\ 
Use of HTTP without SSL/TLS & 6  & 1.0 & 1.0 \\
\hline
Average                     &    & 1.0 & 1.0 \\
\hline
\end{tabular}
\end{table}

%%%%%%%%%%%%%%%%%%%%%%%%%%% Student oracle table Ansible end %%%%%%%%%%%%%%%%%%%%

%%%%%%%%%%%%%%%%%%%%%%%%%%% Student oracle table Chef start  %%%%%%%%%%%%%%%%%%%%

\begin{table}[htb]
\centering
\caption{SLAC's Accuracy for Ansible Security Smell Occurrences Identified Only by Students}
\label{tab-res-slac-chef-student}
\begin{tabular}{  p{4cm}  r  r  r  }
\hline
\textbf{Smell Name} & \textbf{Occurr.}  & \textbf{Precision} & \textbf{Recall}  \\
\hline
Hard-coded secret           & 5  &  1.0  &  1.0 \\
Suspicious comment          & 9  &  1.0  &  1.0  \\
Use of HTTP without SSL/TLS & 11 &  1.0  &  1.0 \\
Use of weak crypto. algo.   & 1  &  1.0  &  1.0 \\
\hline
Average                     &    &  1.0  &  1.0  \\
\hline
\end{tabular}
\end{table}

%%%%%%%%%%%%%%%%%%%%%%%%%%% Student oracle table Chef end  %%%%%%%%%%%%%%%%%%%%

%%%%%%%%%%%%%%%%%%%%%%%%%%%%%%%% REBUTTAL ADDITIONAL ORACLE DATASETS %%%%%%%%%%%%%%%%%%%%%%%%%

\subsubsection{Evaluation of SLAC in Round\#2}
\label{slic-eval-round2}

We describe the oracle dataset construction and SLAC's evaluation for the oracle dataset in Round\#2.  

\textbf{Round\#2-Oracle Dataset for Ansible and Chef}: We use a rater who volunteered for constructing the oracle dataset in Round\#2. The rater is a third year PhD student at Tennessee Tech University, with a three year experience in software security that included experience in studying vulnerabilities and security bug reports. Similar to round\#1, the first author performed additional inspection of the 100 scripts used in round\#2. 

As shown in Tables~\ref{tab-res-slic-ansi-accu-round2} and~\ref{tab-res-slic-chef-accu-round2}, the rater identify 42 and 55 occurrences of security smells respectively for Ansible and Chef scripts. The first author did not find any security smell instances missed by the rater.   

\textbf{Round\#2-Performance of SLAC for Ansible and Chef Oracle Dataset}: We provide SLAC's evaluation performance for Ansible and Chef respectively, in Tables~\ref{tab-res-slic-ansi-accu-round2} and~\ref{tab-res-slic-chef-accu-round2} for Round\#2. Evaluation results of SLAC for the oracle datasets is consistent with the evaluation results in Round\#1. For Ansible we observe decrease in average precision, but not for average recall. For Chef the average precision and recall is same as in round\#1.  

\begin{table}[htb]
\centering
\caption{SLAC's Accuracy for the Ansible Oracle Dataset (Round\#2)}
\label{tab-res-slic-ansi-accu-round2}
\begin{tabular}{ p{4.5cm} r r r }
\hline
\textbf{Smell Name} & \textbf{Occurr.} & \textbf{Precision} & \textbf{Recall} \\ 
\hline
Empty password              & 2  & 1.0 & 1.0 \\ 
Hard-coded secret           & 18 & 0.89& 1.0 \\
No Integrity Check          & 8  & 0.75& 1.0 \\ 
Suspicious comment          & 10 & 1.0 & 1.0 \\ 
Use of HTTP without SSL/TLS & 4  & 1.0 & 1.0 \\
\hline
No smell                    & 75 & 1.0 & 1.0 \\ 
\hline
Average &                        & 0.9 & 1.0 \\
\hline
\end{tabular}
\end{table}

\begin{table}[htb]
\centering
\caption{SLAC's Accuracy for the Chef Oracle Dataset (Round\#2)}
\label{tab-res-slic-chef-accu-round2}
\begin{tabular}{  p{4.5cm}  r  r  r  }
\hline
\textbf{Smell Name} & \textbf{Occurr.}  & \textbf{Precision} & \textbf{Recall}  \\
\hline
Admin by default            & 5  &  1.0  &  1.0 \\
Hard-coded secret           & 10 &  0.75 &  0.75 \\
Suspicious comment          & 20 &  1.0  &  1.0  \\
Unrestricted IP address     & 6  &  1.0  &  1.0 \\
Use of HTTP without SSL/TLS & 7  &  1.0  &  1.0 \\
Missing default             & 9  &  0.89 &  1.0 \\
\hline
No smell                    & 71 &  1.0  &  0.9 \\
\hline
Average                     &    & 0.9   &  0.9  \\
\hline
\end{tabular}
\end{table}

%%%%%%%%%%%%%%%%%%%%%%%%%%%%%%%% REBUTTAL ADDITIONAL ORACLE DATASETS %%%%%%%%%%%%%%%%%%%%%%%%%

\textbf{Dataset and Tool Availability}: The source code of SLAC and all constructed datasets are available online~\cite{paper:dataset}.

%%%%%%%%%%%%%%%%%%%%%%%%%%%%%%%%%%%%%%%%%%%%%%%%% EMPIRICAL STUDY METHODOLOGY  %%%%%%%%%%%%%%%%%%%%%%%%%%%%%%%%%
%%%%%%%%%%%%%%%%%%%%%%%%%%%%%%%%%%%%%%%%%%%%%%%%%%%%%%%%%%%%%%%%%%%%%%%%%%%%%%%%%%%%%%%%%%%%%%%%%%%%%%%%%%%%%%%%

\section{Empirical Study}
\label{empirical}

Using SLAC, we conduct an empirical study to quantify the prevalence of security smells in Ansible and Chef scripts.

\subsection{Datasets}
\label{meth-ds}
We conduct our empirical study with four datasets: two datasets each for Ansible and Chef scripts. We construct two datasets from repositories maintained by Openstack. The other two datasets are constructed from repositories hosted on GitHub. We select repositories from Openstack because Openstack create cloud-based services, and could be a good source for IaC scripts. We include repositories from GitHub, because IT organizations host their OSS projects on GitHub~\cite{rahul:icse18:seip}~\cite{amrit:icse18:seip}. In contrary to our prior research~\cite{me:icse2019:slic}, we only used Openstack datasets as Openstack have made their Ansible and Chef scripts available for download. Ansible and Chef scripts are not available for other organizations, such as Mozilla and Wikimedia.    

As advocated by prior research~\cite{MunaiahCuration2017}, OSS repositories need to be curated. We apply the following criteria to curate the collected repositories:

\begin{itemize}[leftmargin=*]
\item{\textbf{Criterion-1}: At least 11\% of the files are IaC scripts. Prior research~\cite{JiangAdamsMSR2015} reported that in OSS repositories IaC scripts co-exist with other types of files, such as Makefiles. A repository that contaisn a few IaC scripts may not be sufficient for analysis. They~\cite{JiangAdamsMSR2015} observed a median of 11\% of the files to be IaC scripts. By using a cutoff of 11\%, we assume to collect repositories that contain sufficient amount of IaC scripts for analysis.}
\item{\textbf{Criterion-2}: The repository is not a clone.}
\item{\textbf{Criterion-3}: The repository must have at least two commits per month. We use this criterion to identify repositories with frequent activity. Munaiah et al.~\cite{MunaiahCuration2017} used the threshold of at least two commits per month to determine which repositories have enough software development activity.}
\item{\textbf{Criterion-4}: The repository has at least 10 developers. Our assumption is that the criteria of at least 10 developers may help us to filter out repositories with limited development activity. Previously, researchers have used the cutoff of at least nine developers~\cite{me:swan2018:ci}~\cite{amrit:icse18:seip}.}

\end{itemize}  

We answer RQ2 using 14,253 Ansible and 36,070 Chef scripts, respectively, collected from 365 and 448 repositories. Table~\ref{table-criteria-dataset-all} summarizes how many repositories are filtered out using our criteria. We clone the master branches of these repositories. Summary attributes of the collected repositories are available in Table~\ref{table-dataset-summary}.   

\begin{table*}[]
\centering
\caption{OSS Repositories Satisfying Criteria (Sect.~\ref{meth-ds})}
\label{table-criteria-dataset-all}
\begin{tabular}{ p{5cm}  r  r r r }
\toprule
                                   & \multicolumn{2}{c}{\textbf{Ansible}} & \multicolumn{2}{c}{\textbf{Chef}}  \\
\midrule
                                   & GH & OST & GH & OST            \\
\midrule
\textbf{Initial Repo Count}        & 3,405,303 & 1,253 & 3,405,303 & 1,253   \\
\midrule
Criterion-1 (11\% IaC Scripts)      & 13,768 & 16       & 5,472  & 15     \\
Criterion-2 (Not a Clone)           & 10,017 & 16       & 3,567  & 11   \\
Criterion-3 (Commits/Month $\ge$ 2) & 10,016 & 16       & 3,565  & 11   \\
Criterion-4 (Devs $\ge$ 10)         & 349    & 16       & 438 & 10      \\
\midrule
\textbf{Final Repo Count}          & 349    & 16       & 438 & 10     \\
\bottomrule
\end{tabular}
\end{table*}

\begin{table*}[htb]
\centering
\caption{Summary Attributes of the Datasets}
\label{table-dataset-summary}
\begin{tabular}{  p{3.0cm}  r  r  r  r r r}
\hline
                           & \multicolumn{2}{c}{\textbf{Ansible}} & \multicolumn{2}{c}{\textbf{Chef}}  \\
\hline
\textbf{Attribute}         & \textbf{GH} & \textbf{OST} & \textbf{GH} & \textbf{OST} \\
\hline
Repository Count        & 349       & 16      & 438       & 10         \\
Total File Count        & 498,752   & 4,487   & 126,958   & 2,742    \\
Total Script Count      & 13,152    & 1,101   & 35,132    & 938      \\
Tot. LOC (IaC Scripts)  & 602,982   & 52,239  & 1,981,203 & 63,339  \\
\hline
\end{tabular}
\end{table*}

\subsection{Analysis}

\textbf{Sanity Check}: As reported in Section~\ref{meth-slic-eval}, SLAC has a high precision and recall for the oracle dataset, but it may under-perform for scripts not included in our oracle dataset. We mitigate this limitation by creating sanity check datasets for 100 Ansible and Chef scripts each that are not included in the oracle dataset. We select these 100 Ansible and 100 Chef scripts randomly form the Openstack dataset constructed in Section~\ref{meth-ds}. The first author performs sanity check analysis.

For Ansible we observe 20 scripts to contain at least one security smell. SLAC identifies 45, 1, 36, 2, 16, and 9 occurrences of hard-coded secrets, empty passwords, HTTP without TLS usages, unrestricted IP address bindings, suspicious comments, and no integrity checks. Precision of SLAC for hard-coded secrets, empty passwords, HTTP without TLS usage, unrestricted IP address bindings, suspicious comments, and no integrity checks respectively, is 0.7, 1.0, 1.0, 1.0, 1.0, and 0.7. Recall of SLAC for hard-coded secrets, empty passwords, HTTP without TLS usage, unrestricted IP address bindings, suspicious comments, and no integrity checks respectively, is 1.0, 1.0, 1.0, 1.0, 1.0, and 0.9. 

For Chef we observe 19 scripts to contain at least one security smell. SLAC identifies 26, 38, 4, 9, 2, and 4 occurrences of hard-coded secrets, HTTP without TLS usage, unrestricted IP address bindings, suspicious comments, missing default in case instances, and no integrity checks. Precision of SLAC for hard-coded secrets, HTTP without TLS usage, unrestricted IP address bindings, suspicious comments, missing default in case instances, and no integrity checks respectively, is 0.8, 1.0, 1.0, 1.0, 1.0, and 0.8. Recall of SLAC for hard-coded secrets, HTTP without TLS usages, unrestricted IP address bindings, suspicious comments, missing default in case instances, and no integrity checks respectively, is 1.0, 1.0, 1.0, 1.0, 1.0, and 0.9. 

We observe SLAC to generate false positives, but the recall is $>=0.9$ for all security smell categories. SLAC's detection accuracy provides confidence on identifying security smells in other scripts not included in the oracle dataset.  

%%%%%%%%%%%%%%%%%%%%%%%%%%%%%% METH: ANSWER TO RQ2, RQ3 %%%%%%%%%%%%%%%%%%%%%%%%%%
%%%%%%%%%%%%%%%%%%%%%%%%%%%%%%%%%%%%%%%%%%%%%%%%%%%%%%%%%%%%%%%%%%%%%%%%%%%%%%%%%%
%%%%%%%%%%%%%%%%%%%%%%%%%%%%%%%%%%%%%%%%%%%%%%%%%%%%%%%%%%%%%%%%%%%%%%%%%%%%%%%%%%

\subsubsection{Answer to RQ2: How frequently do security smells occur for Ansible and Chef scripts?}
\label{meth-rq2} 

First, we apply SLAC to determine the security smell occurrences for each script. Second, we calculate two metrics described below:  

\begin{itemize}[leftmargin=*]
\item{\textbf{\textit{Smell Density}}: We use smell density to measure the frequency of a security smell $x$, for every 1000 lines of code (LOC). Our smell density metric is similar to that of prior research that have used defect density~\cite{jss:defect:density}, and is measured using Equation~\ref{equ-meth-smell-density}.

\begin{equation}
\begin{aligned}\label{equ-meth-smell-density}
\text{Smell Density ($x$)} = \\ \frac{\text{Total occurrences of $x$}}{\text{Total line count for all scripts}/1000}
\end{aligned}
\end{equation}

}
\item{\textbf{\textit{Proportion of Scripts (Script\%)}}: We use the metric `Proportion of Scripts' to quantify how many scripts have at least one security smell. This metric refers to the percentage of scripts that contain at least one occurrence of smell $x$.}
\end{itemize} 

%The two metrics characterize the frequency of security smells differently. The smell density metric is more granular, and focuses on the content of a script as measured by how many smells occur for every 1000 LOC. The proportion of scripts metric is less granular and focuses on the existence of at least one of the identified security smells for all scripts. 

\subsubsection{RQ3:  How do practitioners perceive the identified security smell occurrences for Ansible and Chef scripts? }

We gather feedback using bug reports on how practitioners perceive the identified security smells. We apply the following procedure: 

\textbf{First}, we randomly select 500 occurrences of security smells for each of Ansible and Chef scripts. \textbf{Second}, we post a bug report for each occurrence, describing the following items: smell name, brief description, related CWE, and the script where the smell occurred. We explicitly ask if contributors of the repository agrees to fix the smell instances. 
\textbf{Third}, we determine a practitioner to agree with a security smell occurrence if (i) the practitioner replies to the submitted bug report explicitly saying the practitioner agrees, or (ii) the practitioner fixes the security smell occurrence in the specified script by running SLAC on IaC scripts, for which we submitted bug reports. If the security smell does not exist in the script of interest, then we determine the smell to be fixed.

%%%%%%%%%%%%%%%%%%%%%%%%%%%%%%%%%%%%%%%%%%%%%%%%% EMPIRICAL FINDINGS %%%%%%%%%%%%%%%%%%%%%%%%%%%%%%%%%%%%%%%%%%
%%%%%%%%%%%%%%%%%%%%%%%%%%%%%%%%%%%%%%%%%%%%%%%%%%%%%%%%%%%%%%%%%%%%%%%%%%%%%%%%%%%%%%%%%%%%%%%%%%%%%%%%%%%%%%%

\section{Empirical Findings}  
\label{results}

We answer RQ2 and RQ3 in this section.  

%%%%%%%%%%%%%%%%%%%%% ANSWER TO RQ2 %%%%%%%%%%%%%%%%%%%%%%

\subsection{Answer to RQ2: How frequently do security smells occur for Ansible and Chef scripts?}
\label{res-rq2}

We observe our identified security smells to exist across all datasets. For Ansible, in our GitHub and Openstack datasets we observe respectively 25.3\% and 29.6\% of the total scripts to contain at least one of the six identified security smells. For Chef, in our GitHub and Openstack datasets we observe respectively 20.5\% and 30.4\% of the total scripts to contain at least one of the eight identified security smells. A complete breakdown of findings related to RQ2 for Ansible and Chef is presented in Tables~\ref{tab-res-occur-all},~\ref{tab-res-density-all}, and~\ref{tab-res-prop-all} for our datasets.    

\textbf{\textit{Occurrences}}: The occurrences of the security smells are presented in the `Occurrences' column of Table~\ref{tab-res-occur-all} for all datasets. The `Combined' row presents the total smell occurrences. In the case of Ansible scripts, we observe 18,353 occurrences of security smells, and for Chef, we observe 28,247 occurrences of security smells. For Ansible, we identify 15,131 occurrences of hard-coded secrets, of which 55.9\%, 37.0\%, and 7.1\% are respectively, hard-coded keys, user names, and passwords. For Chef, we identify 15,363 occurrences of hard-coded secrets, of which 47.0\%, 8.9\%, and 44.1\% are respectively, hard-coded keys, user names, and passwords. 

Exposing hard-coded secrets, such as hard-coded keys, is not uncommon: Meli et al.~\cite{git:secret:ndss2019} studied secret key exposure in OSS GitHub repositories, and identified 201,642 instances of private keys, which included commonly-used API keys. Meli et al.~\cite{git:secret:ndss2019} reported 85,311 of the identified 201,642 instances of private keys to be Google API keys.       

\begin{table*}[htb]
\centering
\caption{Smell Occurrences for Ansible and Chef scripts}
\label{tab-res-occur-all}
\begin{tabular}{  p{5cm}  r   r  r   r      }
\hline
& \multicolumn{2}{c}{\textbf{Ansible}} & \multicolumn{2}{c}{\textbf{Chef}}   \\
\hline
\textbf{Smell Name} & \textbf{GH} & \textbf{OST}  & \textbf{GH} & \textbf{OST}   \\
\hline
Admin by default            & N/A    & N/A   & 301   & 61      \\
Empty password              & 298    & 3     & N/A   & N/A    \\
Hard-coded secret           & 14,409 & 722   & 14,160& 1,203  \\
Missing default in switch   & N/A    & N/A   & 953   & 68      \\
No integrity check          & 194    & 14    & 2,249 & 132     \\
Suspicious comment          & 1,421  & 138   & 3,029 & 161     \\
Unrestricted IP address     & 129    & 7     & 591   & 19     \\
Use of HTTP without SSL/TLS & 934    & 84    & 4,898 & 326    \\
Use of weak crypto algo.    & N/A    & N/A   & 94    & 2       \\
\hline 
\textbf{Combined}           & 17,385 & 968   & 26,275& 1,972  \\
\hline
\end{tabular}
\end{table*}

\begin{table*}[htb]
\centering
\caption{Smell Density for Ansible and Chef scripts}
\label{tab-res-density-all}
\begin{tabular}{  p{5cm}  r   r  r   r    }
\hline
& \multicolumn{2}{c}{\textbf{Ansible}} & \multicolumn{2}{c}{\textbf{Chef}}   \\
\hline
\textbf{Smell Name} & \textbf{GH} & \textbf{OST}  & \textbf{GH} & \textbf{OST}   \\
\hline
Admin by default            & N/A  & N/A  & 0.1  & 0.9      \\
Empty password              & 0.49 & 0.06 & N/A  & N/A     \\
Hard-coded secret           & 23.9 & 13.8 & 7.1  & 19.0  \\
Missing default in switch   & N/A  & N/A  & 0.5  & 1.0       \\
No integrity check          & 0.3  & 0.2  & 1.1  & 2.1    \\
Suspicious comment          & 2.3  & 2.6  & 1.5  & 2.5    \\
Unrestricted IP address     & 0.2  & 0.1  & 0.3  & 0.3     \\
Use of HTTP without SSL/TLS & 1.5  & 1.6  & 2.4  & 5.1     \\
Use of weak crypto algo.    & N/A  & N/A  & 0.05 & 0.03    \\
\hline 
\textbf{Combined}           & 28.8 & 18.5 & 13.3 & 31.5   \\
\hline
\end{tabular}
\end{table*}

\begin{table*}[htb]
\centering
\caption{Proportion of Scripts With At Least One Smell for Ansible and Chef scripts}
\label{tab-res-prop-all}
\begin{tabular}{  p{5cm}  r   r  r   r     }
\hline
& \multicolumn{2}{c}{\textbf{Ansible}} & \multicolumn{2}{c}{\textbf{Chef}}    \\
\hline
\textbf{Smell Name} & \textbf{GH} & \textbf{OST}  & \textbf{GH} & \textbf{OST}    \\
\hline
Admin by default            & N/A  & N/A  & 0.3  & 2.1   \\
Empty password              & 1.1  & 0.2  & N/A  & N/A   \\
Hard-coded secret           & 19.2 & 22.4 & 6.8  & 15.9  \\
Missing default in switch   & N/A  & N/A  & 2.5  & 6.5   \\
No integrity check          & 1.1  & 1.0  & 3.6  & 3.8   \\
Suspicious comment          & 6.3  & 8.0  & 6.6  & 9.3   \\
Unrestricted IP address     & 0.5  & 0.4  & 1.1  & 1.0   \\
Use of HTTP without SSL/TLS & 3.7  & 3.0  & 4.9  & 6.9   \\
Use of weak crypto algo.    & N/A  & N/A  & 0.2  & 0.1   \\
\hline 
\textbf{Combined}           & 25.3 & 29.6 & 20.5 & 30.4   \\
\hline
\end{tabular}
\end{table*}

%%%%%%%%%SMELL DENSITY %%%%%%%%%

\textbf{\textit{Smell Density}}: In Table~\ref{tab-res-density-all}, we report the smell density for both, Ansible and Chef. The `Combined' row presents the smell density for each dataset when all identified security smell occurrences are considered. For all datasets, we observe the dominant security smell to be `Hard-coded secret'.

\textbf{\textit{Proportion of Scripts (Script\%)}}: In Table~\ref{tab-res-prop-all}, we report the proportion of scripts (Script \%) values for each of the four datasets. The `Combined' row represents the proportion of scripts in which at least one of the identified smells appear.

\subsection{Answer to RQ3: How do practitioners perceive the identified security smell occurrences for Ansible and Chef scripts?}
\label{res-rq4} 

\begin{figure}[htb]
\centering
\includegraphics[scale=0.85]{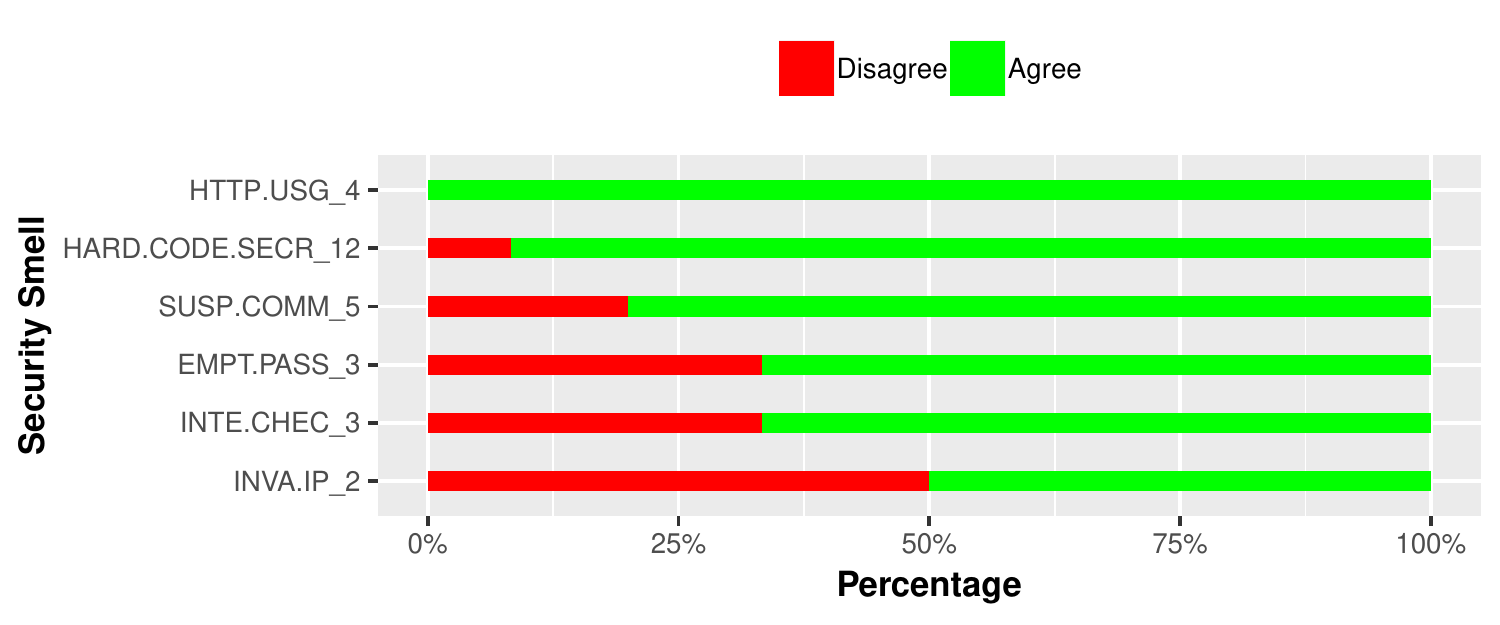}
\caption{Feedback for 29 smell occurrences for Ansible. Practitioners agreed with 82.7\% of the selected smell occurrences.}
\label{fig-res-feedback-ansi}
\end{figure}

\begin{figure}[htb]
\centering
\includegraphics[scale=0.85]{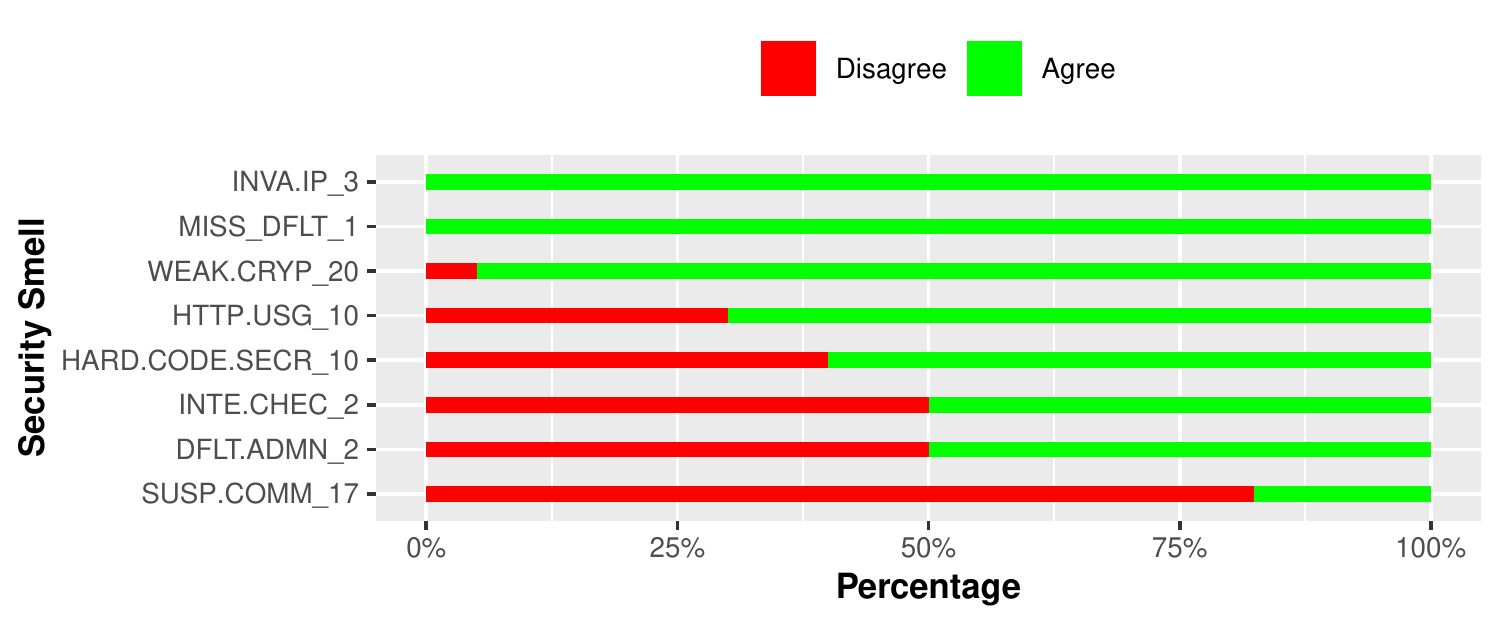}
\caption{Feedback for the 65 smell occurrences for Chef. Practitioners agreed with 63.1\% of the selected smell occurrences.}
\label{fig-res-feedback-chef}
\end{figure}

From 7 and 30 repositories, respectively, we obtain 29 and 65 responses for the submitted 500 Ansible and the 500 Chef security smell occurrences. In the case of Ansible, we observe an agreement of 82.7\% for 29 smell occurrences. For Chef, we observe an agreement of 63.1\% for 65 smell occurrences. The percentage of smells to which practitioners agreed to be fixed for Ansible and Chef is respectively, presented in Figure~\ref{fig-res-feedback-ansi} and~\ref{fig-res-feedback-chef}. In the y-axis each smell name is followed by the occurrence count. For example, according to Figure~\ref{fig-res-feedback-ansi}, for 4 occurrences of `Use of HTTP without SSL/TLS' (HTTP.USG) , we observe 100\% agreement for Ansible scripts. 

We acknowledge that the response rate is 9.4\%, which is low for the submitted bug reports. One possible explanation can be developers might be biased against security smell alerts, as they are typically generated by static analysis tools. Upon submission of the bug reports, developers may have considered the identified security smells as `code smells', and left these bug reports as unresolved. Developers incorrect perceptions on insecure coding is not uncommon: for example, Acar et al.~\cite{acar:secure:insecure:2017} have observed developers bias to perceive their code snippets as secure, even if the code snippets are insecure. 

Another possible explanation can be we have submitted bug reports for repositories that are inactive despite the applying systematic criteria to filter the repositories. For example, for one bug report a practitioner mentioned that the repository `rcbops/ansible-lxc-rpc/' is no longer maintained~\footnote{https://github.com/rcbops/ansible-lxc-rpc/issues/681}. 

Another possible explanation can be lack of actionability: the submitted bug reports do not provide suggestions on how to act on the security smells. As an example, if a hard-coded password appears in an Ansible or Chef script, we do not discuss what techniques should be adopted to repair the smell occurrence in the bug report.

\textbf{\textit{Reasons for Practitioner Agreements}}: Lack of awareness and availability of repair suggestions attributed to why practitioner agreed to security smell instances. We provide examples below: 

\begin{itemize}
    \item{\textit{Awareness of HTTPS availability}: We submitted a bug report for two instances of `HTTP without SSL/TLS'. For both instances a URL was used to download RStudio packages we submitted a bug report~\footnote{https://github.com/elasticluster/elasticluster/issues/634}. In response to the bug report, the practitioner agreed that the smell instance needs to be repaired, and repaired the smell instances~\footnote{https://github.com/elasticluster/elasticluster/commit/a62b8aae6559a3a15fbb724709005caba8cf33e8}. The practitioner also stated why the smell instance was introduced in the first place: ``\textit{In this case, I think it was just me being a bit sloppy: the HTTPS endpoint is available so I should have used that to download RStudio packages from the start}''. 
    } 
    \item{\textit{Awareness on hard-coded secrets}: For an instance of hard-coded user name and hard-coded password for an Ansible script~\footnote{https://github.com/quarkslab/irma/blob/master/ansible/playbooks/group\_vars/all.yml} we submitted a bug report~\footnote{https://github.com/quarkslab/irma/issues/60}. In response, the practitioner acknowledged the presence of the smells. The practitioner also stated what actions he can take to mitigate the security smells: ``\textit{I agree that it [hard-coded secret] could be in an Ansible vault or something dedicated to secret storage.}''.    
    } 
    \item{\textit{Availability of repair suggestions}: For one instance of weak cryptography usage in a Chef script, we submitted a bug report~\footnote{https://github.com/cookbooks/hw-postgresql/issues/1}. Along with submitting the bug report, we also submitted a pull request which replaced MD5 usage with SHA512~\footnote{https://github.com/cookbooks/hw-postgresql/pull/2/commits/66f9841177080988d2af9789f92daa4c0a1b325d}. The pull request was accepted a month later, and presence of the security smell was acknowledged by the practitioner.    
    }
\end{itemize}

\textbf{\textit{Reasons for Practitioner Disagreements}}: For disagreements we observe development context to be an important factor. We provide examples below: 

\begin{itemize}
    \item{\textit{Dependency}: Practitioners may disagree with instances of `use of HTTP without SSL/TLS' for an URL, if the URL refers to a dependency maintained by an external organization, upon which the practitioner or the team has no control upon. For example in a bug report~\footnote{https://github.com/cookbooks/ic-cassandra/issues/2}, we observe a practitioner to disagree with occurrences of `use of HTTP without SSL/TLS' in a Chef script~\footnote{https://github.com/cookbooks/ic-cassandra/blob/master/cookbooks/hadoop\_cluster/recipes/add\_cloudera\_repo.rb}. All of these URLs refer to remote archive hosts maintained by Cloudera~\footnote{https://www.cloudera.com/}, an IT organization that provides cloud utilities. The practitioner disagreed and asked to report this issue to `upstream', i.e. to the project maintainers who manage the URLs. }
    
    \item{\textit{Location of smells}: A hard-coded password may not have security implications for practitioners if the hard-coded password is located in testing code of Chef or Ansible scripts. In one bug report~\footnote{https://github.com/Graylog2/graylog2-cookbook/issues/109} a practitioner stated ``\textit{the code in question is an integration test. The username and password is not used anywhere else so this should be no issue.}''. The practitioners views were echoed for another instance of hard-coded password, which we reported as a bug report~\footnote{https://github.com/chef-cookbooks/docker/issues/1069}. The practitioner also provided suggestions on how we can prioritize inspection: ``\textit{I suggest that the author probably needs to adjust his scanner to not be quite so sensitive when it detects usernames and passwords set in RSpec or Inspec code. Or at least to prompt the person running the script before creating an issue on a repository. Human intervention is likely the best principled action, here.}''. For both bug reports the practitioners assume that hard-coded usernames and passwords in test code is not relevant as the hard-coded password will never be used in production system. one possible limitation to such assumption is that practitioners are only considering their own development context, and not realizing how another practitioner, not experienced in IaC, may perceive use of these security smells as an acceptable practice. As documented in GitHub issues for bug resolution, developers have strong perceptions about bugs identified by research tools on whether they are `important' or not. For example, developers of Z3 strongly disagreed on a bug reported by researchers because the identified bug is ``\textit{asinine}''. Furthermore, the developer adds ``\textit{As someone who uses Z3/Boolector/STP/CVC4 1000s of times a day, I would much rather that issue trackers such as these are full-up with issues that real users find, than the ones you derive.}''~\footnote{https://github.com/Z3Prover/z3/issues/4461}.         
    }
\end{itemize}

\section{Discussion}
\label{discussion}

We discuss implications of our paper as following:

\subsection{Towards Actionable Detection and Repair: Lessons Learned}
\label{disc-actionability}

Practitioner responses from the submitted bug reports provide signals on how we can make SLAC more actionable with respect to detection. We have learned that practitioners do not consider hard-coded user names and passwords in testing scripts as relevant. Toolsmiths can take this observation into account and tune future security smell detection tools accordingly. We also learned that URL instances related to `use of HTTP without TLS' might not be relevant if an HTTPS URL exists in the first place. For example, in the case of generating a security smell alert, for $http://archive.cloudera.com/debian/archive.key$, SLAC could have been adjusted to check for the availability of a secure HTTP endpoint for the URL. 

Along with submitting bug reports for the detected static analysis alerts, automated pull requests can be generated that will include repair suggestions for the detected security smell instances. Practitioners might be more receptive to a security smell instance, if the alert notification also accompanies suggestions on how to repair. For example, toolsmiths can create tools that will generate automated pull requests, which will show how to repair a security smell instance.

\subsection{Additional Contributions Compared to Our Prior Research}  
\label{disc-contribs} 

As discussed in Section~\ref{bg-differences}, the three IaC languages, namely, Ansible, Chef, and Puppet are different to each other with respect to execution order, perceived codebase maintenance, requiring additional agent software installation, style and syntax. The above-mentioned differences merit a systematic investigation of the categorization and quantification of security smells for Ansible and Chef scripts. From a practitioner perspective, if a team only uses Ansible scripts, then the catalog of security smells for Puppet scripts and the tool to detect security smells from our prior work may not be relevant. Similarly, for practitioners using Chef, the security smell catalog and tool for Puppet or Ansible might not be relevant. We have noticed anecdotal evidence related to this: upon completion of our prior work, we reached out to practitioners for feedback. One feedback was `` This is practical. Does the tool work for Ansible?''. Our replication study addresses the needs of practitioners who use Ansible or Chef. 

Schwarz et al.~\cite{schwarz:code:smell:iac} pursued similar efforts for Chef's code maintenance smells research. They~\cite{schwarz:code:smell:iac} replicated Sharma et al.~\cite{SharmaPuppet2016}'s research on Puppet code maintenance smells for Chef scripts, and observed Puppet’s code smells to appear for Chef as well. Schwarz et al.~\cite{schwarz:code:smell:iac}'s paper is an example how replication can benefit the researcher community to advance the science of IaC script quality. We too have followed the footsteps of Schwarz et al.~\cite{schwarz:code:smell:iac}, and replicated our prior research for Ansible and Chef.    

In short, the differences in contributions between our prior work and this paper is following:
\begin{itemize}

\item{A list of security smells for Ansible and Chef scripts that includes two categories not reported in prior work\cite{me:icse2019:slic}}; 

\item{An evaluation of frequency of security smells occur in Ansible and Chef scripts. As a result of this evaluation we have created a benchmark how frequently security smells appear for Ansible and Chef. Till date such benchmark is missing. Frequency of identified security smells for Ansible and Chef scripts can be used as a measuring stick by practitioners and researchers alike};

\item{A detailed discussion on how practitioner responses from bug reports can drive actionable detection and repair of Ansible and Chef security smells. In our prior work we did not discuss how the practitioners responses in bug reports can guide tools for actionable detection and repair};

\item{An empirically-validated tool (SLAC) that automatically detects occurrences of security smells for Ansible, and Chef scripts. The tool that we constructed as part of prior work will not work for Ansible and Chef scripts. The `Parser' component of SLAC is different to that of `SLIC' that we built in our prior work. The `Rule Engine' component of SLAC is different to that of SLIC, as unlike Puppet, which uses attributes, Ansible and Chef respectively, uses `Keys' and `Properties'}; and 

\item{A detailed discussion of differences between the three IaC languages, Ansible, Chef, and Puppet. In our prior work, we provided background on Puppet scripts only, and did not discuss the differences between Ansible, Chef, and Puppet.}
\end{itemize}

\subsection{Differences in Security Smell Occurrences for Ansible, Chef, and Puppet Scripts}  
\label{disc-diff} 

Our identified security smells for Ansible and Chef overlap with Puppet. The security smells that are common across all three languages are: hard-coded secret, suspicious comment, unrestricted IP address, and use of HTTP Without SSL/TLS. Security smells identified for Puppet are also applicable for Chef and Ansible, which provides further validation to our prior research findings~\cite{me:icse2019:slic}. We also identify additional security smells namely, `Missing Default in Case' and `No Integrity Check', which was not previously identified by Rahman et al.~\cite{me:icse2019:slic}. One possible explanation can be related to rater bias: in our prior work, we used one rater to identify security smells in Puppet scripts. The rater might have missed instances of `Missing Default in Case' and `No Integrity Check'. Another possible explanation can be the set of scripts the rater used for inspection in prior work~\cite{me:icse2019:slic}. Perhaps, those set of scripts were carefully developed by developers who are aware of the security consequences related to the new categories.

Despite differences in frequency of security smells across datasets and languages, we observe the proportion of scripts to contain at least one smell varies between 20.5\% and 32.9\%. Our findings indicate that some IaC scripts, regardless of their languages may include operations that make those scripts susceptible to security smells. Our finding is congruent with Rahman and Williams' observations~\cite{me:icst2018:iac}: they observed defective Puppet scripts to contain certain operations: operations related to filesystem, infrastructure provisioning, and user account management. Based on our findings and prior observations from Rahman and Williams~\cite{me:icst2018:iac} we conjecture that similar to defective scripts, IaC scripts with security smells may also include certain operations that distinguishes them from scripts with no security smells.   

Our results related to Ansible and Chef overlap that with prior research~\cite{me:icse2019:slic}. Despite the overlap, our results have implications for practitioners, toolsmiths, and educators. The fact that our research results related to Chef and Ansible overlap the findings for Puppet highlights a lack of awareness related to security for IaC. Regardless of what IaC language is being used, certain security smells, such as hard-coded secrets are dominant. Practitioners who are using Ansible, Chef, or Puppet scripts should be aware of the security consequences. Toolsmiths can build upon our tools, SLIC~\cite{me:icse2019:slic} and SLAC to detect security smells in other IaC languages, such as Terraform. Educators who are teaching DevOps and configuration management should discuss the security implications of security smells in IaC scripts. The commonality of the identified security smells across the three programming languages provides evidence related to robustness of our prior work conducted only on Puppet~\cite{me:icse2019:slic}. Endres et al.~\cite{endres:swe:diversity} suggested that research results confirmed by diverse data sources help advance scientific research in the field of software engineering.

\subsection{On the Value of Replication for IaC Research}  
\label{disc-replication} 

Presented findings in our paper is an example on why IaC-related research should be replicated for other programming languages. We have identified two new security smell categories that were not reported in prior research. For IaC-related research typically researchers have relied on Chef and Puppet~\cite{me:ist:sms:2018}. However, structural and semantic differences exist between IaC-related programming languages, and the IaC community may benefit from replication studies, which can identify differences and similarities in research conclusions across multiple programming languages. For example, Rahman and Williams~\cite{me:ist2019:code:properties}'s work on identifying source code properties can be replicated for a larger set of scripts developed in other languages, such as Ansible. As another example, Hummer et al.~\cite{Hummer:idem:chef:2013}'s research on Chef idempotency can be replicated for Ansible and Puppet scripts. In multiple blog posts~\footnote{https://www.simplilearn.com/ansible-vs-puppet-the-key-differences-to-know-article}~\footnote{https://www.devopsgroup.com/blog/puppet-vs-ansible/}~\footnote{https://blog.gruntwork.io/why-we-use-terraform-and-not-chef-puppet-ansible-saltstack-or-cloudformation-7989dad2865c}, practitioners have mentioned how one language can be different to another with respect to syntax, scalability, and configuration management philosophy~\footnote{https://www.upguard.com/articles/ansible-puppet}. The domain of IaC research can benefit from replication studies where IaC scripts written in two or more languages can be used to confirm or negate research hypotheses.

\subsection{Mitigation Strategies}
\label{disc-smells}

\hspace{0.3cm}\textbf{Admin by default}: Practitioners can follow the recommendations from Saltzer and Schroeder~\cite{Saltzer:Schroeder:1975} to create user accounts that have the minimum possible security privilege and use that account as default. 

\textbf{Empty password}: The use of strong passwords can mitigate appearance of empty passwords in Ansible and Chef scripts.  

\textbf{Hard-coded secret}: We provide two suggestions: \textit{first}, scan IaC scripts to search for hard-coded secrets using tools such as CredScan~\footnote{https://secdevtools.azurewebsites.net/helpcredscan.html} and SLAC. \textit{Second}, use tools such as Ansible/AWS~\footnote{https://github.com/ansible/awx} and Vault~\footnote{https://www.vaultproject.io/} to store secrets. 

\textbf{Missing default in case statement}: We advise programmers to always add a default `else' block so that unexpected input does not trigger events, which can expose information about the system.

\textbf{No integrity check}: As IaC scripts are used to download and install packages and repositories at scale, we advise practitioners to always check downloaded content by computing hashes of the content or checking with GPG signatures.

\textbf{Suspicious comment}: We acknowledge that in OSS development, programmers may be introducing suspicious comments to facilitate collaborative development and to provide clues on why the corresponding code changes are made~\cite{ICSE:TODO:Storey}. Based on our findings we advocate for creating explicit guidelines on what pieces of information to store in comments, and strictly follow those guidelines through code review. For example, if a programmer submits code changes where a comment contains any of the patterns mentioned for suspicious comments in Table~\ref{tab-res-slic-kw}, the submitted code changes will not be accepted.

\textbf{Unrestricted IP address}: To mitigate this smell, we advise programmers to allocate their IP addresses systematically based on which services and resources need to be provisioned. For example, incoming and outgoing connections for a database containing sensitive information can be restricted to a certain IP address and port.

\textbf{Use of HTTP without SSL/TLS}: We advocate companies to adopt the HTTP with SSL/TLS by leveraging resources provided by tool vendors, such as MySQL~\footnote{https://dev.mysql.com/doc/refman/5.7/en/encrypted-connections.html} and Apache~\footnote{https://httpd.apache.org/docs/2.4/ssl/ssl\_howto.html}. We advocate for better documentation and tool support so that programmers do not abandon the process of setting up HTTP with SSL/TLS.

\textbf{Use of weak cryptography algorithms}: We advise programmers to use cryptography algorithms recommended by the National Institute of Standards and Technology~\cite{nist:crypt:rec} to mitigate this smell. For example, `MD5' usages should be replaced by `SHA256' or `SHA512'.

\subsection{Future Work}
\label{disc-future}

From Section~\ref{res-rq2}, answers to RQ2 indicate that not all IaC scripts include security smells. Researchers can build upon our findings to explore which characteristics correlate with IaC scripts with security smells. If certain characteristics correlate with scripts that have smells, then programmers can prioritize their inspection efforts for scripts that exhibit those characteristics. Researchers can investigate how semantics and dynamic analysis of scripts can help in efficient smell detection. Researchers can also investigate what remediation strategies can be adopted that facilitate better actionability and repair of security smells identified by SLAC. As our detection accuracy results indicate, SLAC generates false positives, which can motivate future work to detect security smells with high precision.

We have not quantified lifetime of security smells for Ansible and Chef scripts. Quantifying the lifetime of security smells for Ansible and Chef scripts is an excellent idea, which will require significant change in the design of SLAC. Currently, SLAC detects the presence of security smells in Ansible and Chef scripts. For lifetime detection SLAC should be expanded to handle (i) code snippets where smell appears, (ii) obtaining each version of all 50,323 scripts over 9 years, and (iii) using heuristics to compare code snippets across 9 years that include security smells. Researchers can investigate the lifetime of Ansible and Chef scripts in future. 

%%%%%%%%%%%%%%%%%%%%%%%%%%%%%%%%%%%%%%%%%%%%%%%%% THREATS  %%%%%%%%%%%%%%%%%%%%%%%%%%%%%%%%%%%%%%%%%%%%%%%%%%%%
%%%%%%%%%%%%%%%%%%%%%%%%%%%%%%%%%%%%%%%%%%%%%%%%%%%%%%%%%%%%%%%%%%%%%%%%%%%%%%%%%%%%%%%%%%%%%%%%%%%%%%%%%%%%%%%

\section{Threats to Validity}
\label{threats}

In this section, we discuss the limitations of our paper: 

\textbf{Conclusion Validity}: The derived security smells and their association with CWEs are subject to the rater judgment. During the security smell derivation process the first author was involved, who also derived the security smells for Puppet scripts~\cite{me:icse2019:slic}. The first author's bias can influence the smell derivation process for Ansible and Chef scripts. We account for this limitation by using another rater, second author of the paper, who is experienced in software security.

The oracle datasets were constructed by the raters. The construction process is susceptible to subjectivity, as the raters' judgment influences appearance of a certain security smell. We mitigate this limitation by allocating at least two raters for each script. We have used graduate students to construct oracle datasets. But as reported in Section~\ref{meth-slic-eval}, students miss security smell occurrences. We mitigate this limitation by using the first author who identified security smell instances missed by the graduate students. However, in the process, bias inherent in the first author's judgement can influence the construction of the oracle dataset. We mitigate this limitation by constructing another oracle dataset with a volunteer rater.   

We use certain thresholds to curate repositories based on observations reported in prior research~\cite{JiangAdamsMSR2015}~\cite{MunaiahCuration2017}~\cite{amrit:icse18:seip}. Our selection thresholds can be limiting. For example, a repository may contain sufficient amount of Ansible or Chef scripts, but maintained by one practitioner. Such repositories even though active, will be excluded in our analysis based on criteria mention in Section~\ref{empirical}.  

\textbf{Internal Validity}: We acknowledge that other security smells may exist for both Ansible and Chef. We mitigate this threat by manually analyzing 1,101 Ansible and 855 Chef scripts for security smells. In the future, we aim to investigate if more security smells exist.

The detection accuracy of SLAC depends on the constructed rules that we have provided in Tables~\ref{tab-res-ansi-rules} and~\ref{tab-res-chef-rules}. We acknowledge that the constructed rules are susceptible of generating false positives and false negatives. Accuracy of SLAC is dependent on the string patterns used in Table~\ref{tab-res-slic-kw}.  

\textbf{External Validity}: Our findings are subject to external validity, as our findings may not be generalizable. We observe how security smells are subject to practitioner interpretation, and thus the relevance of security smells may vary from one practitioner to another. Also, our scripts are collected from the OSS domain, and not from proprietary sources. We conduct our investigation with two languages, Ansible and Chef. Investigation of other languages used in IaC, such as Terraform, can reveal new categories of security smells. Also, reported detection accuracy for SLAC is limited to the two oracle datasets and the sanity check dataset.

%%%%%%%%%%%%%%%%%%%%%%%%%%%%%%%%%%%%%%%%%%%%%%%%% CONCLUSION  %%%%%%%%%%%%%%%%%%%%%%%%%%%%%%%%%%%%%%%%%%%%%%%%%
%%%%%%%%%%%%%%%%%%%%%%%%%%%%%%%%%%%%%%%%%%%%%%%%%%%%%%%%%%%%%%%%%%%%%%%%%%%%%%%%%%%%%%%%%%%%%%%%%%%%%%%%%%%%%%%

\section{Conclusion}
\label{conclusion}

IaC is the practice of using automated scripting to provision computing environments by applying recommended software engineering practices, such as version control and testing. Security smells are recurring coding patterns in IaC scripts that are indicative of security weakness and can potentially lead to security breaches. By applying open coding on 1,101 Ansible and 855 Chef scripts, we identified six and eight security smells respectively, for Ansible and Chef. The security smells that are common across all three languages are: hard-coded secret, suspicious comment, unrestricted IP address, and use of HTTP Without SSL/TLS. 

Next, we construct a static analysis tool called SLAC using which we analyzed 50,323 Ansible and Chef scripts. We identify 46,600 security smells by analyzing 50,323 scripts, which included 7,849 hard-coded passwords. Based on smell density, we observed the most dominant smell to be `Hard-coded secret'. We observe security smells to be prevalent in Ansible and Chef scripts. We recommend practitioners to rigorously inspect the presence of the identified security smells through code review and by using automated static analysis tools for IaC scripts. We hope our paper will facilitate further security-related research in the domain of IaC scripts.

%%%%%%%%%%%%%%%%%%%%%%%%%%%% ACKNOWLEDGEMENT %%%%%%%%%%%%%%%%%%%%%%%%%%
\begin{acks}
We thank the RealSearch group at NC State University and the anonymous reviewers for their valuable feedback. Our research was partially funded by the NSA's Science of Security Lablet at NC State University. We also thank Farzana Ahamed Bhuiyan of Tennessee Technological University for help in expanding the oracle dataset for SLAC's evaluation.   
\end{acks}

%\balance

\bibliographystyle{ACM-Reference-Format}
\bibliography{slic}             

\end{document}